\documentclass[aps,nofootinbib,superscriptaddress,floatfix,prd,twocolumn,letterpaper]{revtex4} %eqsecnum,
\usepackage{graphicx,amsmath,amssymb,color,bm}
\usepackage{hyperref}
\usepackage{epsfig}
\usepackage{epstopdf}

\def\be{\begin{equation}}
\def\ee{\end{equation}}
\def\bea{\begin{eqnarray}}
\def\eea{\end{eqnarray}}
\newcommand{\vs}{\nonumber\\}
\def\ba#1\ea{\begin{align}#1\end{align}}
\def\Msunh{\,h^{-1}M_{\odot}}
\def\iMpch{\,h\,{\rm Mpc}^{-1}}
\def\Mpch{\,h^{-1}\,{\rm Mpc}}

\newcommand{\refeq}[1]{Eq.~(\ref{eq:#1})}

\newcommand{\reffig}[1]{Fig.~\ref{fig:#1}}          
          
\newcommand{\refsec}[1]{Sec.~\ref{sec:#1}}          
\newcommand{\refapp}[1]{App.~\ref{app:#1}}
\newcommand{\reftab}[1]{Table~\ref{tab:#1}}

\newcommand{\bfem}[1]{\textbf{\emph{#1}}}

\renewcommand{\v}[1]{\bm{#1}}

\newcommand{\vx}{\v{x}}
\newcommand{\vy}{\v{y}}
\newcommand{\xfl}{\v{x}_{\rm fl}}

\newcommand{\vv}{\v{v}}

\newcommand{\vk}{\v{k}}
\newcommand{\vq}{\v{q}}

\def\lapl{\nabla^2}
\def\lapln{\nabla^{2n}}
\newcommand{\<}{\left\langle}
\renewcommand{\>}{\right\rangle}

\renewcommand{\d}{\delta}

\def\s{\sigma}

\newcommand{\vnhat}{\v{\hat{n}}}

\newcommand{\vn}{\boldsymbol{\nabla}}
\newcommand{\Om}{\Omega_m}

\def\B{\mathcal{B}}

\def\sigmaT{\sigma_\text{T}}

\newcommand{\comment}[1]{}

\def\cH{\mathcal{H}}

\def\vnhat{\hat{\v{n}}}

\def\zre{z_\text{re}}
\def\knl{k_\text{NL}}

\begin{document}

\title{Imprints of Reionization in Galaxy Clustering}

\author{Fabian Schmidt}
\affiliation{Max-Planck-Institut f\"ur Astrophysik, Karl-Schwarzschild-Str.~1, 85748~Garching, Germany}

\author{Florian Beutler}
\affiliation{Institute of Cosmology \& Gravitation, Dennis Sciama Building, University of Portsmouth, Portsmouth, PO1 3FX, UK}

\begin{abstract}
  Reionization, the only phase transition in the Universe since recombination, is a key event in the cosmic history of baryonic matter. We derive, in the context of the large-scale bias expansion, the imprints of the epoch of reionization in the large-scale distribution of galaxies, and identify two contributions of particular importance. First, the Compton scattering of CMB photons off the free electrons lead to a drag force on the baryon fluid. This drag induces a relative velocity between baryons and CDM which is of the same order of magnitude as the primordially-induced relative velocity, and enters in the evolution of the relative velocity as calculated by Boltzmann codes. This leads to a unique contribution to galaxy bias involving the matter velocity squared. The second important effect is a modulation of the galaxy density by the ionizing radiation field through radiative-transfer effects, which is captured in the bias expansion by so-called higher-derivative terms. We constrain both of these imprints using the power spectrum of the BOSS DR12 galaxy sample. While they do not lead to a shift in the baryon acoustic oscillation scale, including these terms is important for unbiased cosmology constraints from the shape of the galaxy power spectrum.
\end{abstract}

\date{\today}

\maketitle

\section{Introduction}

The evolution of baryonic matter in the Universe is characterized by
two phase transitions: the first, recombination, happened at redshift
$z_*\approx 1100$, after which baryons formed a neutral gas and were
decoupled from radiation. The second phase transition, reionization,
happened at some point between $\zre \sim 6$ and $\zre \sim 20$. After
reionization, baryons were essentially fully ionized again. The history of reionization is not nearly as well determined as that of recombination (see \cite{barkana/loeb:01,zaroubi:12} for reviews). The reason for
this is that the radiation sources that brought about this phase transition
are surmised to be stars, explosive events such as supernovae, or
accreting black holes; that is, highly nonlinear objects whose formation is
incompletely understood. On the other hand, the progenitors of galaxies
observed in the low-redshift Universe were actively forming around the
epoch of reionization, which is thus expected to have strong effects
on these progenitors. For example, reionization is one of the possible
explanations for the suppression of the stellar-mass-to-halo-mass ratio $M_*/M_h$ observed
in low-mass galaxies today \cite{efstathiou:92,barkana/loeb:99}.

While it is thus well established that reionization can strongly
affect the mean number density of tracers (for example, galaxies identified at
fixed stellar mass), much less study has been devoted to the question of
how reionization affects the large-scale \emph{clustering} of observed
galaxies. In this paper, we address this question, focusing on scales
much larger than the nonlinear scale, $\knl \sim 0.3\iMpch$ or $R_{\rm NL} \sim 20\Mpch$ today, where perturbation theory applies. On these scales, the
clustering of galaxies can be related to that of the total matter distribution
through a perturbative bias expansion (see \cite{biasreview} for a comprehensive
overview). Thus, the key question we attempt to address is:
\begin{itemize}
\item[] Which additional contributions does reionization add to the bias expansion?
\end{itemize}

In particular, we will study three distinct physical effects:
\begin{itemize}
\item Compton drag
\item pressure forces
\item radiative-transfer effects.
\end{itemize}
As we will see, the second and third contributions are captured by terms in the
bias expansion, so-called higher-derivative terms, that are well known and present already even when all non-gravitational effects are neglected. Reionization will however affect the magnitude of these contributions; in particular, the radiative-transfer effects could lead to significant effects on comoving scales of order $50\Mpch$. On the other hand, the effect of Compton drag leads to contributions that are \emph{distinct from all previously considered terms in the bias expansion of galaxies}.

Our considerations involve two steps. First, we consider the physical mechanisms by which reionization can influence galaxy formation, by way of the evolution
of the large-scale baryon density (\refsec{deriv}). Then, allowing
for the galaxy density to depend on the baryon density along its entire
past history, and within a finite region, we write down the contributions
to the general perturbative galaxy bias expansion (\refsec{bias}). 

This result then allows us to constrain the amplitude of the Compton drag
and radiative-transfer effect on galaxies using the data release 12 (DR12) sample from the Baryon Oscillations
Spectroscopic Survey (BOSS). These constraints are presented in \refsec{constraints}. We conclude in \refsec{concl}.

We spell out the complete expression for the 1-loop galaxy power spectrum which is being fitted to the data in \refapp{Pkmodel}. 
App.~B provides an estimate of the adiabatic decaying mode induced by pre-recombination plasma oscillations, which has so far not been considered in essentially all studies on galaxy bias. We show that it is very small even during the dark ages before reionization, and, thus, that it can indeed be neglected.

We assume a flat $\Lambda$CDM fiducial cosmology with $\Omega_m=0.31$, $\Omega_bh^2=0.022$, $h=0.676$, $\sigma_8=0.824$, $n_s=0.96$ and $\sum m_{\nu} = 0.06\,$eV, which is the fiducial cosmology used in the BOSS data analysis. 
Numerical results have been obtained from the Boltzmann code CAMB \cite{camb}, in the version of January 2016. 

\section{Reionization effects on the baryon distribution}
\label{sec:deriv}

In this section, we consider the effects of reionization on the large-scale
baryon density and velocity. As a starting point, we will work to linear order in perturbation theory, as the effects are small on large scales, and the nonlinear scale $R_\text{NL}$, i.e. the spatial scale within which the fractional rms density contrast is order one, is very small at high redshifts. The incorporation of nonlinear
evolution will then be considered in the next section.

\subsection{Compton drag}
\label{sec:drag}

On large scales, both baryons and CDM can be described as pressureless fluids which are coupled by gravity (we will consider baryonic pressure below). Then, at linear order in perturbations, their velocities $\vv_b, \vv_c$ each follow the linearized Euler equation,
\ba
\v{v}_b' + \cH \v{v}_b =\:& -\vn \Phi \vs
\v{v}_c' + \cH \v{v}_c =\:& -\vn \Phi \,,
\label{eq:eul1}
\ea
where primes denote derivatives w.r.t. conformal time $\tau$, $\cH = a'/a$, and the gravitational potential $\Phi$ obeys the Poisson equation,
\be
\lapl\Phi = \frac32 \Om(\tau) \cH^2 \d_m\,,
\label{eq:Poisson}
\ee
where $\Om(\tau)$ is the ratio of the mean matter density to the critical density, and $\d_m$ is the fractional perturbation in the matter density. This is in turn given by the fractional density perturbations $\d_b, \d_c$ in baryons and CDM, respectively, through
\be
\d_m = f_b \d_b + (1-f_b) \d_c\,,
\ee
where $f_b = \Omega_b/\Omega_c$ is the ratio of mean baryon and mean CDM densities.
Correspondingly, the matter velocity is given by\footnote{Note that at nonlinear order, various definitions of the coarse-grained matter velocity can be chosen. This is however irrelevant for our results, since the differences are absorbed by the bias parameters appearing in the nonlinear bias expansion.}
\be
\v{v} \equiv \vv_m = f_b \v{v}_b + (1-f_b) \v{v}_c\,.
\ee

\refeq{eul1} states that the relative velocity $\vv_r$ between baryons and CDM is not sourced, and obeys
\be
\v{v}_r + \cH \v{v}_r = 0\,,\quad\mbox{where}\quad
\v{v}_r \equiv \v{v}_b - \v{v}_c\,.
\ee
The source-free nature remains true also at nonlinear order.  Thus, in terms of the relative velocity $\vv_r(\tau_{\rm in}) \equiv \vv_{bc}$ at some initial time $\tau_\text{in}$, we have 
\ba
\vv_r(\vx,\tau) =\:& \vv_{bc}(\vx) \frac{a_\text{in}}{a(\tau)}\,,
\label{eq:vbcprim}
\ea
and the relative velocity decays as $1/a$. Here, we have adopted the commonly used notation $\vv_{bc}$ for the initial conditions of the relative velocity $\vv_r$. Before recombination, the baryons were tightly coupled to photons. This coupling led to an effective force term on the r.h.s. of the Euler equation \refeq{eul1} for baryons (see below), which induces a relative velocity. After recombination, this velocity $\vv_{bc}$, which we will refer to as the ``primordial contribution'' from now on, decays following \refeq{vbcprim}. The impact of this relative velocity on large-scale structure was first pointed out by \cite{tseliakhovich/hirata:2010} (but see also \cite{shoji/komatsu,somogyi/smith:2010}) and has been studied extensively since then
\cite{dalal/etal:2010,yoo/dalal/seljak,tseliakhovich/barkana/hirata,Visbal/etal:12,popa/etal,yoo/seljak,bernardeau/vdr/vernizzi,slepian/eisenstein,lewandowski/perko/senatore,blazek/etal:15,schmidt:2016b,ahn}.  

Let us now consider what happens during and after reionization. As soon as the ionization fraction is significant, the gas is weakly coupled to the freely streaming relic photons of the cosmic microwave background (CMB)
through Compton scattering.  Specifically, the photons scatter off the
electrons, which in turn are bound to the nuclei through Coulomb forces.  
Due to the significantly diluted baryon density as well as redshifted 
relic radiation at the epoch of reionization, this coupling is much weaker than that before recombination.  
Nevertheless, we will see that it is not negligible in terms of the baryon-CDM
relative velocity.  
In this subsection, we will keep the speed of light $c$ explicit.  Note that this
means that the gravitational potential $\Phi$ has units of $c^2$.  

\begin{figure*}[t]
\centering
\includegraphics[trim=1.6cm 2.5cm 1.2cm 1.8cm,clip,width=0.8\textwidth]{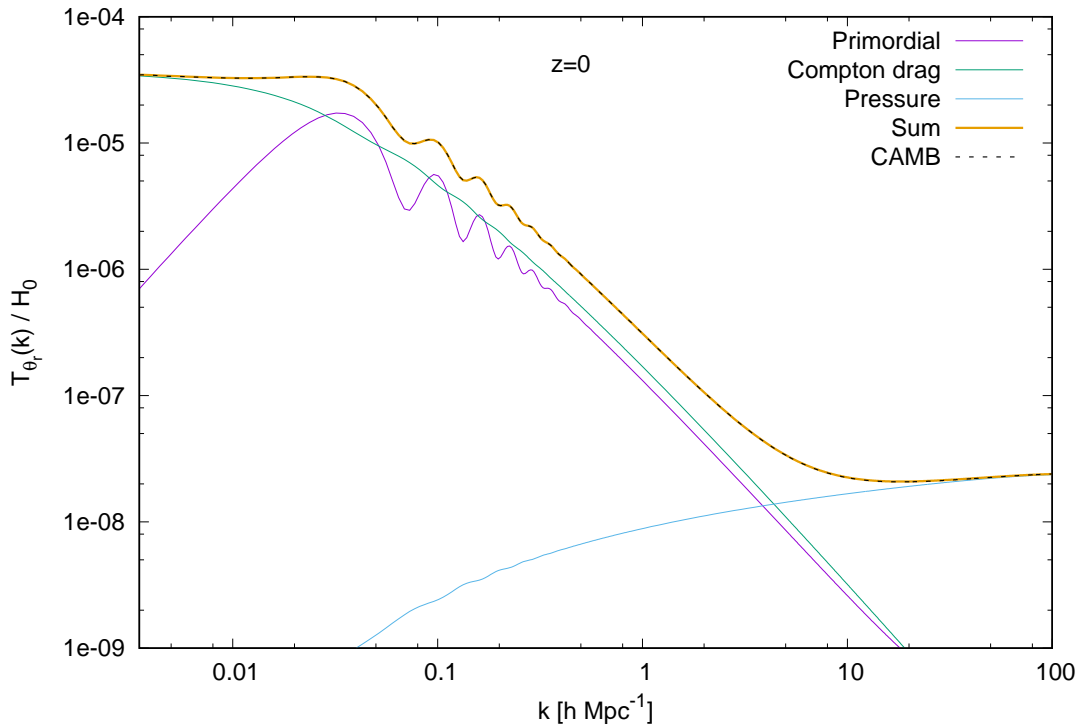}
\caption{Transfer function for the divergence of the baryon-CDM relative velocity $\theta_r$ (divided by $H_0$ to be dimensionless) at $z=0$. We show the primordial contribution, obtained by using a pre-reionization transfer function output from CAMB evolved to $z=0$, as well as the Compton-drag and pressure contributions as estimated in the text [\refeq{thetabcdrag} and \refeq{thetabcpres}; we have adjusted the value of $\alpha_0$ from the approximate estimate \refeq{alpha} by $\sim$20\%].
  Also shown is the total transfer function, as well as the output from CAMB at $z=0$, which agrees well. 
All transfer functions are normalized so that $T_m(k\to 0,z=0)=1$.  
\label{fig:Tk}}
\end{figure*}
Each electron elastically scatters photons at a 
rate of $\sigmaT n_\gamma c$, where $\sigmaT$ is the Thomson cross section, and $n_\gamma$ is the number density of photons.  
The momentum transfer in each scattering is approximately $k_{\rm B}T_\gamma$, where $T_\gamma = 2.73 (1+z)\,{\rm K}$ is the radiation temperature, since the photon 
momentum is much smaller than that of the electron.  Integrating over
the photon energy, the total momentum transfer rate on a single electron
is $\sigmaT u_\gamma$, where
\be
u_\gamma = \frac{8\pi^5}{15 (h c)^3} (k_{\rm B} T_\gamma)^4
\ee
is the energy density of radiation.  However, since the radiation field
is close to isotropic (radiation density perturbations are at the level of
$10^{-4}$ and negligible here), an electron at rest in the CMB rest frame
experiences no net force.  If the electron moves with velocity
$\v{v}_{\gamma e}$ relative to the CMB rest frame, it experiences a
drag due to the fact that there is a radiation dipole given by $(\v{v}_{\gamma e}/c) T_\gamma$ 
in the electron rest frame.  Thus, the drag force on the electron is
\be
\v{F}_e = - \frac{\v{v}_{\gamma e}}c \sigmaT u_\gamma\,.
\ee
Multiplying this by the electron density yields the force density
$\v{f}_e = n_e \v{F}_e$, which, via the Coulomb coupling between electrons
and nuclei, contributes to the r.h.s. of the 
baryon Euler equation through $a \v{f}_e/\rho_b$, where $\rho_b$ is the baryon density (the factor of $a$ arises because we have written the Euler equation in terms of $\partial/\partial\tau$).  Choosing the CMB rest frame as defining the global coordinate system, as is usually done, we have $\v{v}_{\gamma e} = \v{v}_b$.  
Adding this to \refeq{eul1}, we have
\ba
\v{v}_b' + \cH \v{v}_b =\:& -\vn \Phi - x_e \alpha \cH\, \v{v}_b
\,,
\label{eq:eulb2}
\ea
where $x_e(\tau)$ is the electron ionization fraction, and the dimensionless
function $\alpha$ is given by
\ba
\alpha(\tau) =\:& a(\tau) \frac{\sigmaT u_\gamma(\tau)}{Y_e m_p c \cH(\tau)} 
= \alpha_0 (1+z)^4 E^{-1}(z)\,, \vs
\alpha_0 \approx\:& 1.61 h^{-1}\times 10^{-6} 
\,,
\label{eq:alpha}
\ea
and $Y_e\approx 1.08$ is the electron molecular weight, $m_p$ is the proton
mass, and $E(z) = H(z)/H_0$ is the scaled Hubble rate.  
\begin{figure*}[t]
\centering
\includegraphics[trim=1.6cm 2.5cm 1.2cm 1.8cm,clip,width=0.8\textwidth]{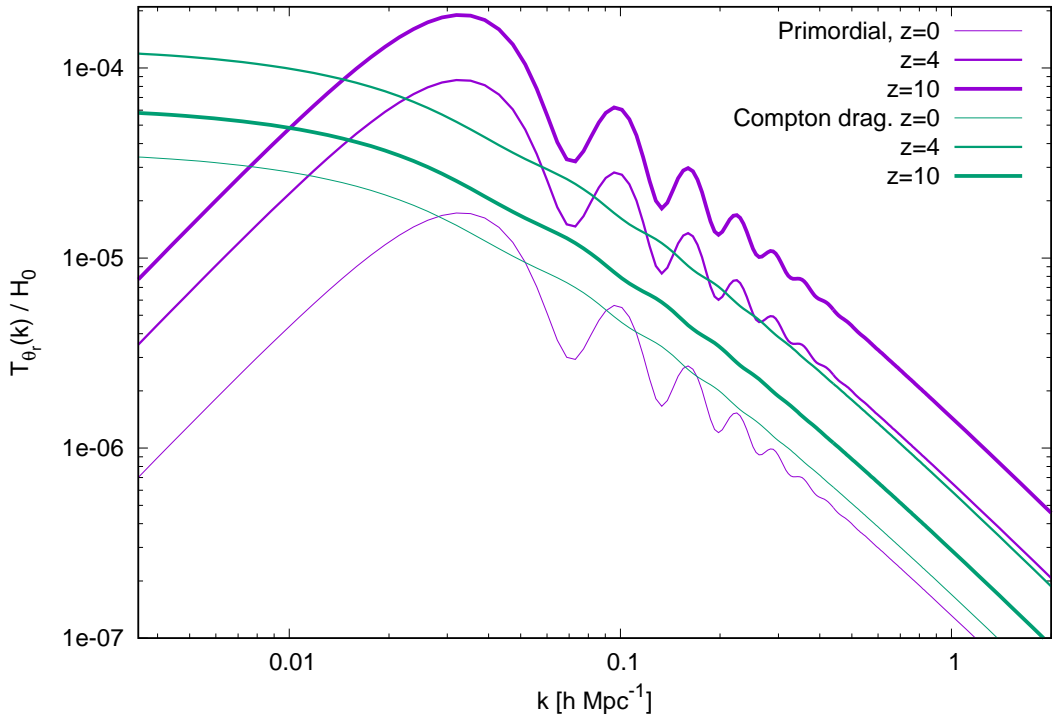}
\caption{
  Primordial and Compton-drag contributions to the transfer function of the baryon-CDM relative velocity divergence at different redshifts. The different scaling with redshift of these physically distinct contributions is evident. In particular, the Compton drag behaves non-monotonically, due to an interplay of the buildup of the relative velocity and redshifting of the background radiation [see \refeq{vbcdragest}].
\label{fig:Tkz}}
\end{figure*}

We finally obtain the equation of motion for the baryon-CDM relative
velocity, at linear order in perturbations, in the presence of Compton drag:
\ba
\v{v}_r' + \cH \v{v}_r =\:& - x_e \alpha \cH \v{v}\,.
\label{eq:vbcdrag}
\ea
Here, we have approximated the baryon velocity
with the total matter velocity $\vv$. This approximation is sufficient, since the difference between
$\v{v}_b$ and $\v{v}$ is suppressed by 2--3 orders
of magnitude, while, as we will see, the contribution on the r.h.s. of
\refeq{vbcdrag} is itself already highly suppressed. Moreover, since we are working at linear order in perturbations, the contribution from \refeq{vbcdrag} simply adds to the primordial relative velocity of \refeq{vbcprim}. We can then immediately integrate \refeq{vbcdrag} to obtain
\ba
& \vv_r(\vx,\tau) = C_\text{drag}(\tau) \vv(\vx,\tau) + \vv_{bc}(\vx) \frac{a_\text{in}}{a(\tau)}\,,
\label{eq:vbcdrag2} \\
& C_\text{drag}(\tau) \equiv \frac{\alpha_0}{a^2 E(\tau) f D(\tau)} \int_{\ln a_\text{in}}^{\ln a(\tau)}\!\!\!\!\!\! d\ln a'\:
x_e(a') a'^{-2} f D(a')\,,
\label{eq:Cdragdef}
\ea
where $D(\tau)$ is the linear growth factor while $f \equiv dD/d\ln a$ is the growth rate. 
Taking the divergence of \refeq{vbcdrag2}, we obtain
\ba
\theta_r(\vx,\tau) =\:& C_\text{drag}(\tau) \theta(\vx,\tau)
+ \theta_{bc}(\vx) \frac{a_\text{in}}{a(\tau)}\,,
\label{eq:thetabcdrag} 
\ea
where $\theta_{bc} = \vn\cdot\vv_{bc}$. 
Thus, Compton drag leads to a source term for the relative velocity divergence between baryons and CDM which is given by the matter velocity divergence $\theta = -\delta_m'$, which is in turn proportional to the matter density perturbation at linear order. 
\reffig{Tk} shows this contribution to the baryon-CDM relative velocity at redshift $z=0$, where we have assumed instantaneous reionization at $\zre \approx 11.8$ (as determined by CAMB from the given optical depth). Here, the transfer function is defined as
\be
T_{\theta_r}(k,z) = \frac52 \Omega_{m0} \frac{H_0^2}{c^2 D(z=0)} \frac{\theta_r(\vk,z)}{k^2\mathcal{R}(\vk)}\,,
\label{eq:Tdef}
\ee
where $\Omega_{m0}$ is the matter density parameter today, the growth factor $D(z)$ is normalized to $D(z) = 1/(1+z)$ during matter domination, and $\mathcal{R}$ is the primordial superhorizon curvature perturbation. We similarly define the transfer function for other quantities. Note that $T_m(k) \equiv T_{\d_m}(k,z=0) \to 1$ for $k \ll 0.01\iMpch$ using this normalization. We thus see that, at $z=0$, the baryon-CDM relative velocity is suppressed by more than four orders of magnitude compared to the velocity of the adiabatic growing mode (using that $T_\theta(k,z=0)/H_0 \sim T_m(k)$).

We also show the primordial contribution in \reffig{Tk}, obtained by scaling the pre-reionization CAMB prediction for $\theta_r$ to $z=0$ using the $1/a$ scaling [\refeq{vbcprim}]. In fact, this contribution is a factor of a few smaller than the Compton-drag contribution at $z=0$. It shows strong baryon acoustic oscillation (BAO) features, which are absent in the Compton-drag contribution, being proportional to the matter velocity. The primordial relative velocity is suppressed on scales $k \lesssim 0.03 \iMpch$, corresponding to scales outside the sound horizon at recombination. This suppression is not present in the Compton-drag contribution, since all perturbations within the comoving horizon at the given redshift contribute to the local CMB dipole. 
Finally, we see that the sum of the two contributions (thick yellow line) agrees extremely well with the transfer function for $\theta_r$ output by CAMB at $z=0$, for $k \lesssim 2\iMpch$. This shows that these two contributions indeed completely determine the baryon-CDM relative velocity on large scales. The strongly rising contribution on small scales will be considered in the next section. 

\reffig{Tkz} illustrates the redshift evolution of the primordial and Compton-drag contributions. While the primordial contribution strictly decays as $1/a$, the Compton-drag contribution shows a peak at intermediate redshifts. This can be understood from \refeq{Cdragdef}. 
In matter domination and assuming instantaneous reionization at $a_{\rm re} = (1+z_{\rm re})^{-1}$, \refeq{Cdragdef} allows for an analytical solution, yielding 
\be
\frac{v_r}{v}\Big|_\text{drag} = C_\text{drag}(a) \stackrel{\text{EdS}}{=} \alpha_0\, a^{-3/2} \left[ a_{\rm re}^{-1} - a^{-1} \right]\,,
\label{eq:vbcdragest}
\ee
for $a > a_{\rm re}$ ($C_\text{drag}$ is trivially zero otherwise). 
For $z_{\rm re} \approx 11$, the largest value is obtained at $z \sim 6-7$, with $C_\text{drag} \sim 2.5\times 10^{-4}$. At higher redshifts, not enough time has elapsed since reionization for the Compton drag to have an effect, while at lower redshifts, the decay of peculiar velocities together with the rapid redshifting of the momentum flux in the CMB radiation suppress the drag-induced relative velocity. The magnitude of the numerical result at $z=4$ shown in \reffig{Tkz} also matches the expectation from \refeq{vbcdragest}. We will use this estimate for our order-of-magnitude estimates below. It is also worth pointing out that, after hydrogen reionization is complete, the reionization of HeII \cite{reimers/etal:1997} supplies additional electrons, and will thus contribute to the Compton drag.

The fact that the primordial and Compton-drag contributions to the baryon-CDM relative velocity originate from different epochs and scale differently with time shows that we have to include them separately in the galaxy bias expansion. We will turn to this in \refsec{bias}.

\subsection{Pressure forces}
\label{sec:pres}

\reffig{Tk} shows that, while the prediction of the low-redshift baryon-CDM relative velocity from CAMB is accurately represented as due to primordial and post-reionization Compton drag, this no longer holds on very small scales. We now show that the small-scale contribution to $\vv_r$ is due to pressure. Pressure perturbations add a term $-\rho_b^{-1} \vn p$
to the baryon Euler equation \refeq{eul1}, where $p$ is the pressure.  Note that a homogeneous pressure component has no dynamical effect. Defining the sound speed through
\be
\delta p = c_s^2 \d \rho_b\,,
\ee
we then obtain at linear order
\ba
\v{v}_b' + \cH \v{v}_b =\:& -\vn \Phi - a(\tau) c_s^2(\tau) \vn \d_b\,,
\label{eq:vbcpres}
\ea
where we did not include the Compton-drag term. Since we are working at linear order in perturbations, it suffices to consider these contributions separately. 
Again, the factor of $a(\tau)$ is due to our use of conformal time. 
We see that, due to the additional derivative, these
pressure perturbations become relevant on small scales, depending
on the sound speed.  For reference, the sound speed of atomic hydrogen
($m=m_p$, $\gamma =5/3$) is given by
\be
\frac{c_s}{c} = 4\times 10^{-7}\left(\frac{T}{1\:{\rm K}}\right)^{1/2}\,.
\ee
Subtracting the dark matter velocity and taking the divergence of \refeq{vbcpres}, we have
\ba
& \theta_r(\vx,\tau)\Big|_\text{pressure} = - C_p(\tau) H_0^{-1} \nabla^2\d_m(\vx,\tau)\,,
\label{eq:thetabcpres} \\
& C_p(\tau) \equiv \frac1{a(\tau)D(\tau)} \int_{\ln a_\text{in}}^{\ln a(\tau)} d\ln a'\:
a' E^{-1}(a') D(a') \vs
&\hspace*{3cm} \times\left[c_s^2(a') - c_{s,\text{CDM eff}}^2(a') \right]\,.
\label{eq:Cpdef}
\ea
Here, we have again approximated $\d_b \approx \d_m$, as the difference is at the percent level on large scales \cite{barkana/loeb:11,schmidt:2016b}.
The second term in \refeq{Cpdef} contains the effective sound speed $c_{s,\text{CDM eff}}$ of dark matter. While cold dark matter is pressureless microscopically, an effective sound speed (along with anisotropic stress) is present when coarse-graining the dark matter velocity field on finite scales \cite{Baumann,carrasco/etal:12,hertzberg:2014}, which accounts for the breakdown of the fluid approximation on nonlinear scales. 

\refeq{thetabcpres} shows that the pressure contribution to the baryon-CDM relative velocity is captured by a so-called higher-derivative term, $\lapl\d_m$ (see Sec.~2.6 of \cite{biasreview} for an overview). 
\reffig{Tk} shows that this contribution describes perfectly the small-scale contributions to the baryon-CDM relative velocity as calculated by CAMB. Here, we have estimated $C_p$ using the adiabatic scaling for the gas temperature, $T \propto (1+z)^2$, yielding $c_s^2(a) = c_{s,0}^2 (1+z)^2$, and matched $c_{s,0}$ to the CAMB result. This yields
\be
\frac{c_{s,0}}{c} \approx 6\times 10^{-7}\,,
\ee
which corresponds to a temperature today of $\sim 3\,{\rm K}$.  Given the absence of
heat sources in CAMB, this is roughly the expected magnitude of the gas temperature today. Of course, the actual evolution of the relative velocity between baryons and CDM on these very small scales, which are in the fully nonlinear regime at low redshifts, is very different from that predicted by CAMB.

In reality, the first stars photo-heat the gas to much higher temperatures, $T\sim 10^4\,{\rm K}$, which correspondingly increase the pressure forces. Nevertheless, we will see in the next section that the effect of pressure is still expected to be subdominant compared to the effects of ionizing radiation on large scales.

\subsection{Radiative-transfer effects}
\label{sec:RT}

The epoch of reionization had a significant effect on the thermal state
of the gas, and thus strongly affected the process of galaxy formation (see \cite{barkana/loeb:01,zaroubi:12} for reviews). While before
reionization, the rate of gas accretion onto halos was set by the cooling
time of the gas, the rapid heating of the gas by reionization, to temperatures of order $10^4\,$K, led to photo-evaporation of gas from low-mass halos \cite{barkana/loeb:99}, and halted the accretion onto halos with masses below the now greatly increased Jeans mass (as $M_{\rm J} \propto T^{3/2}$). For reference, the Jeans mass for an ideal monatomic gas of $T=10^4\,{\rm K}$ is $M_{\rm J} \approx 1.7\times 10^7 \Msunh$, corresponding to a comoving Jeans scale of $\lambda_{\rm J} \approx 0.08\Mpch$. 
This in turn suppressed the formation of galaxies in such low-mass halos
\cite{efstathiou:92}. On the other hand, these
effects are not expected to have strongly affected the formation of galaxies in
halos that were much more massive than $M_{\rm J}$ around reionization.

Now, for the purposes of describing galaxy clustering, the key question is
whether this suppression happened in a spatially homogeneous fashion, or
was modulated by large-scale perturbations.  On small scales, 
reionization most likely occurred in a highly inhomogeneous fashion. The first
sources of ionizing radiation propagated ionization fronts into the neutral
medium, forming Stromgren spheres of ionized, heated gas (if the sources emitted a
significant fraction of X-rays, such as expected for active galactic nuclei and microquasars,
then the gas outside of the ionization front might also have been heated appreciably).
One thus expects that, for halos not much more massive than $M_{\rm J}$,  
the probability of forming a galaxy within a given halo will
depend on the local ionizing background, and thus on the presence of such an ionizing source in the neighborhood of
the proto-galaxy during reionization. 
On the other hand, as soon as the
Stromgren spheres overlapped and the Universe was largely ionized, the ionizing
background is determined by an average over a volume of order the mean free path
of photons, and so became approximately homogeneous. In this regime, the modulation of the galaxy density by radiative-transfer effects is suppressed.

In order to make progress, we now assume that the radiative-transfer effects
can be captured by allowing for the abundance of galaxies $n_g(\vx,z)$, more precisely,
the progenitors of galaxies observed at lower redshifts, to depend
on the local, spherically averaged flux $J_\text{ion}$ of ionizing radiation (see \cite{haardt/madau:96} for a detailed discussion),
\be
J_{\rm ion}(\vx,z) \equiv \int_{912\text{\AA}/c}^\infty \frac{d\nu}{h\nu} \int d^2\vnhat\: I_\nu(\vx,z,\vnhat, \nu)\,,
\label{eq:Jion}
\ee
where $I_\nu$ is the specific intensity of the local radiation field incident on the proto-galaxy, and $912\,$\AA~is the wavelength corresponding to the Lyman limit. We thus write
\ba
& n_g(\vx, z \sim \zre)\Big|_\text{rad. transfer} = F[J_\text{ion}(\vx, z)]\,,
\label{eq:ngRT}
\ea
where $F[J_\text{ion}]$ is a (most likely nonlinear) function. Ignoring any scattering, the flux of ionizing radiation is given by
\ba
J_\text{ion}(\vx,z) = \int d^3\vy\: \frac{\epsilon_\text{ion}(\vx+\vy,z)}{y^2} \exp[-\hat\tau(\vx,\vy,z)]\,,
\label{eq:J2}
\ea
where $\epsilon_\text{ion}$ is the emissivity per unit volume of ionizing radiation (see e.g. \cite{becker/bolton:13} for recent observational constraints), understood as the effective emissivity that enters the intergalactic medium (IGM), i.e. after the escape fraction is applied.  Further, $\hat\tau(\vx,\vy,z)$ is the optical depth of ionizing radiation at redshift $z$ between points $\vx$ and $\vx+\vy$,
\ba
\hat\tau(\vx,\vy,z) =\:& |\vy| \int_0^1 ds\, n_\text{HI}(\vx+s\vy,z) \sigma_{\rm bf}\,,
\ea
where $n_\text{HI}$ is the number density of neutral hydrogen, and
$\sigma_{\rm bf}$ is the bound-free cross section.  All of $\epsilon_\text{ion}, \hat\tau, \sigma_{\rm bf}$ are understood to be averages over frequency [cf. \refeq{Jion}] weighted by the source spectra. We have also approximated the lightcone integrations as spatial integrations at fixed time, since the mean free path of ionizing photons is much smaller than the horizon during reionization.  
We thus see that the integral in \refeq{J2} extends over a region whose size is of order the local mean free path of ionizing radiation, which is given by $\lambda_\text{ion} = 1/(n_\text{HI}\s_{\rm bf})$. This means that the abundance of galaxies depends in detail on the distribution of sources within $\lambda_\text{ion}$, as well as the ionization fraction which determines $n_\text{HI}$. 

Let us then consider what this implies for a large-scale perturbation in the source density with wavelength much greater than $\lambda_\text{ion}$; analogous reasoning applies to variations in the ionization fraction, which themselves are controlled by the emissivity.  Writing $\epsilon_\text{ion}(\vx,z) = \bar \epsilon_\text{ion}(z) [1 + \d_\epsilon(\vx,z)]$, we have
\ba
J_\text{ion}(\vx) =\:& \bar \epsilon_\text{ion} \int \frac{d^3\vy}{y^2}\: [1 + \d_\epsilon(\vx+\vy)] \exp[-|\vy|/\lambda_\text{ion} ] 
\,,
\ea
where we have dropped the redshift argument for clarity, and assumed that the mean free path is spatially uniform. If the perturbation in $\epsilon_\text{ion}$ has a wavelength much longer than $\lambda_\text{ion}$, then we can perform a formal Taylor series of $\d_\epsilon(\vx+\vy)$ around $\vx$ to obtain
\ba
J_\text{ion}(\vx) =\:& \bar \epsilon_\text{ion} \int \frac{d^3\vy}{y^2}\: \left[1 + \d_\epsilon(\vx) + \frac16 \lapl\d_\epsilon(\vx) y^2 + \cdots\right] \vs
& \qquad \times \exp[-|\vy|/\lambda_\text{ion} ] \vs
=\:& \bar J_\text{ion} \left[1 + \d_\epsilon(\vx)
  + \frac13 \lambda_\text{ion}^2 \lapl\d_\epsilon(\vx) + \cdots \right]\,.
\label{eq:Iionexp}
\ea
We have dropped the term at linear order in $\vy$, $\vy\cdot\vn\d_\epsilon$, as we are interested in the case where we average over small-scale fluctuations. Since there is no other preferred direction, the term $\propto \vn\d_\epsilon$ has to vanish after this averaging. 
In the second line, we have performed the now trivial integrations over $\vy$, and related the integral over the mean emissivity to the mean intensity $\bar J_\text{ion}$ of ionizing radiation. Clearly, this is modulated by the perturbation in emissivity. In the limit of small mean free path of ionizing radiation $\lambda_\text{ion}\to 0$, this modulation is local in the perturbation to the emissivity. The leading correction from the finite distance which ionizing radiation travels is captured by the last term, which involves $\lapl\d_\epsilon$, and is multiplied by a coefficient of order $\lambda_\text{ion}^2$.

This shows that the radiative-transfer effects on the clustering of proto-galaxies can, on large scales, be captured by the same higher-derivative terms which are induced by pressure [cf. \refeq{thetabcpres}]. However, instead of the sound horizon, the amplitude of these terms is now set by the mean-free path of ionizing radiation $\lambda_\text{ion}$. Here we have assumed that the fluctuations in the emissivity, loosely speaking the source density, of ionizing radiation themselves trace the matter density on scales of order $\lambda_\text{ion}$. This is likely to be a good physical approximation. A similar reasoning applies to the modulation of the neutral hydrogen density $n_\text{HI}$, which determines the local modulation of $\lambda_\text{ion}$. While distinguishing between the modulation of the ionizing background and the modulation of the mean-free path is important for understanding the physics of reionization and the high-redshift IGM, from the point of view of large-scale galaxy clustering, their effects are both captured by the higher-derivative expansion and thus phenomenologically the same.

Carrying the expansion in \refeq{Iionexp} to higher order, we obtain higher powers of derivatives, such as $\lambda_\text{ion}^4 \nabla^4\d_\epsilon$,
as well as nonlinear terms in $\d_\epsilon$. These contributions to the galaxy density are only suppressed if one considers the clustering of galaxies on scales $r$ much larger than $\lambda_\text{ion}$. If $r \sim \lambda_\text{ion}$, the perturbative treatment of the radiative-transfer effect breaks down. Thus, for galaxies whose progenitors were significantly affected by photo-ionization and heating, $\lambda_\text{ion}$ could set a lower limit on the scales on which their clustering statistics can be described perturbatively. Our task thus is to estimate the size of $\lambda_\text{ion}$. 

Since $n_\text{HI}$ decreases as a function of time, both due to the cosmological expansion
and the increasing ionization fraction, one expects $\lambda_\text{ion}$ to increase. This is confirmed by observations from quasar spectra. Ref.~\cite{worseck/etal:14} report values in comoving units of $\lambda_\text{ion} = 44 \Mpch -  86\Mpch$ at redshifts $z=5.2 - 4.6$, which were evaluated for Lyman-limit photons of 912\AA. Similarly, Ref.~\cite{becker/etal:15} find significant variations in the optical depth, averaged along the line of sight over bins of comoving width $50\Mpch$, of the Ly$\alpha$ forest at $z\sim 5-6$ (note that they attribute these variations to fluctuations in the mean-free path rather than the UV emissivity; however, as discussed above, the relevant scale for the modulation of galaxy clustering is still the mean $\lambda_\text{ion}$).  
Over this redshift range, the bulk of the IGM is already ionized, i.e. the epoch of reionization is essentially completed. If photo-evaporation and heating effects are still relevant at these relatively low redshifts for the galaxy sample under consideration, they could lead to a significant imprint in the clustering of galaxies at comoving scales in the same range. 
If on the other hand the effects of the ionizing radiation on galaxy formation cease soon after reionization is completed, then the relevant value of $\lambda_\text{ion}$ is likely to be significantly smaller. 

We reiterate that for galaxies residing in massive halos ($M_h \gg M_{\rm J} \sim 10^7 \Msunh$), the radiative-transfer effects are not expected to be relevant, as the heating of the gas does not significantly affect their accretion of gas from the IGM. One thus expects that any signature of these radiative-transfer effects will be a strong function of parent halo mass. 
Note that higher-derivative contributions are also present for dark matter halos even when considering gravity only. In that case however, the relevant scale is the Lagrangian radius $(3M/4\pi\bar\rho_m)^{1/3}$
\cite{elia/ludlow/porciani:2012,
  paranjape/sefusatti/etal:2013,biagetti/chan/etal:2014,
  angulo/etal:2015,
  fujita/etal:2016}.  

\section{Contributions to the galaxy bias expansion}
\label{sec:bias}

We now turn to deriving the leading contributions to the general galaxy
bias expansion induced by the three physical effects discussed above:
Compton-drag, pressure forces, and radiative-transfer effects. The bias expansion is defined as an expression for the galaxy density field of the form \cite{biasreview}
\be
\d_g(\vx,\tau) \equiv \frac{n_g(\vx,\tau)}{\bar n_g(\tau)} - 1 = \sum_O b_O(\tau) O(\vx,\tau)\,.
\label{eq:biasgen}
\ee
Here, $n_g$ is the local number density of galaxies, while $\bar n_g$ is the mean at fixed conformal time or redshift. $O$ stands for operators constructed out of the density field, tidal field, baryon-CDM relative velocity, and so on. Each operator has a bias coefficient $b_O$ which is specific to any given galaxy sample and only depends on time, and in general needs to be determined by fitting to galaxy statistics, as we will do in the following. Strictly speaking, the operators $O$ need to be renormalized in order to obtain a meaningful perturbative expansion \cite{mcdonald:2006,PBSpaper,assassi/etal,senatore:2014,MSZ}. For the purposes of this paper, this is a technical detail and not essential for the developments that follow.

The operators $O$ can be further classified by the number of fields they contain (for example, $O=\delta_m$ is first order, while $\delta_m^2$ is second order, and so on), and by the number of spatial derivatives involved. Any operator that involves spatial derivatives on the matter density or tidal field is defined as a \emph{higher-derivative} operator. The significance is that these terms only become relevant on small scales. As already mentioned above, the pressure and radiative-transfer contributions enter as such higher-derivative terms, which have been considered previously. On the other hand,
Compton drag leads to unique contributions that are new to the bias expansion \refeq{biasgen}.
In this section, we revert to units where the speed of light $c=1$. Hence, $H_0 \approx (3000\Mpch)^{-1}$. We also denote $\d \equiv \d_m$, since $\d_b,\,\d_c$ will no longer appear in what follows.  

\subsection{Compton drag}
\label{sec:bias:drag}

The baryon-CDM relative velocity $\vv_r$ is a local observable, and hence
must be included in the general perturbative galaxy bias expansion.

\subsubsection{Linear order}

We begin with the expansion at linear order, where there is only one contribution from the relative velocity, $\theta_r$ \cite{schmidt:2016b}, given in \refeq{thetabcdrag}:
\ba
\theta_r(\vx,\tau) =\:& C_\text{drag}(\tau) \theta(\vx,\tau) + \theta_{bc}(\vx) \frac{a_\text{in}}{a(\tau)}\,.
\label{eq:thetabctot}
\ea
In general, galaxies observed at some time $\tau$ can depend on the relative velocity at any time during their formation history. Thus, fully generally at linear order, the contribution of a relative velocity divergence to the bias relation is given by an integration over the past fluid trajectory,
\be
\d_g(\vx,\tau) \supset \int_0^\tau d\tau'\: F_{\theta_r}(\tau, \tau') \theta_r(\xfl(\tau'),\tau')\,,
\label{eq:thbias}
\ee
where $F_{\theta_r}(\tau,\tau')$ is a kernel specific to the galaxy sample
considered, and $\xfl(\tau')$ denotes the fluid trajectory (geodesic) leading to the
spacetime location $(\vx,\tau)$; at lowest order in perturbations, this is simply $\xfl = \vx=$~const.
As is clear from \refeq{thetabctot},
there are two large-scale contributions to $\theta_r$ with different
time dependences: the recombination contribution $\theta_{bc} a_\text{in}/a$, and the Compton-drag contribution $C_\text{drag} \theta = - C_\text{drag} \cH f \d$, where we have used the linear continuity equation for matter.  We can then
formally perform the time integral in \refeq{thbias}, leading to
\be
\d_g(\vx,\tau) \supset b_\theta^{bc}(\tau) \theta_{bc}(\vx) \frac{a_\text{in}}{a(\tau)} + b_{\text{drag},\theta}(\tau) \d(\vx,\tau)\,,
\label{eq:thbias2}
\ee
where the bias parameters $b_\theta^{bc}$ and $b_{\text{drag},\theta}$ are given by integrals of the kernel $F_{\theta_r}$ against the specific time dependences of each term. Usually, these parameters need to be determined by a fit to the data.

The first term is just the primordial contribution considered in \cite{blazek/etal:15,schmidt:2016b}.  The second term is induced by Compton drag.  However, it is identical
in shape to the ordinary linear bias contribution $b_1 \d$, and thus already taken into account
when allowing for $b_1$ to be a free parameter.  We thus see that, at linear
order, the bias expansion is unchanged from that described in \cite{schmidt:2016b}.  Note that, if a post-reionization transfer function output of a Boltzmann code is used to calculate $\theta_{bc}$, then $b_\theta^{bc}$ captures a mixture of the primordial baryon-CDM velocity and Compton-drag effects.  Thus, if one is interested in the former
effect, a pre-reionization transfer function output should be used to
calculate $\theta_{bc}$, as was done in \cite{beutler/seljak/vlah}.  

\subsubsection{Nonlinear order}

We now turn to the nonlinear bias expansion. As argued in \cite{schmidt:2016b}, the displacement between the baryon and CDM fluids induced by the primordial relative velocity is very small, much smaller than the scales amenable to a perturbative description. This still holds even when the Compton-drag contribution is included. Thus, it is sufficient to treat both baryons and CDM as traveling along the same fluid trajectory $\xfl(\tau)$. The effect of the relative displacement is then captured perturbatively by higher-derivative terms \cite{schmidt:2016b}.

Following \cite{schmidt:2016b}, the terms introduced into the general nonlinear bias expansion by the baryon-CDM relative velocity consist of all combinations of $v_r^i$ and $\partial_i v_r^j$ with the terms that appear in the standard bias expansion which considers only the adiabatic growing mode of the baryon-CDM system \cite{senatore:2014,MSZ}. Ref.~\cite{schmidt:2016b} provides a complete list of these terms for the primordial contribution $v_{bc}$. We now consider which additional terms are added by the Compton-drag contribution. First, for each term which involves $v_r^i$ without a derivative, we obtain the corresponding terms by separating $\vv_r$ into $\vv_{bc}$ (the primordial contribution) and $\vv$ (the Compton-drag contribution). For example, the term $v_r^2$, which was first pointed out by \cite{tseliakhovich/hirata:2010}, leads to
\be
v_r^2 \longrightarrow v^2,\  \v{v}\cdot\v{v}_{bc}\,\  (v_{bc})^2\,,
\ee
which corresponds to the quadratic order effect of Compton drag (as argued
above, the linear order contribution is degenerate with the ordinary density
bias), the coupling between Compton drag and the primordially produced
relative velocity, and the previously considered primordial relative velocity squared. Note that, in the absence of non-gravitational forces, operators in the bias expansion which involve the matter velocity without any derivatives are forbidden by the equivalence principle. However, the local CMB radiation corresponds to a locally identifiable preferred frame, with respect to which the velocity is defined. 

Next, consider the terms involving $\partial_i v_r^j$. The Compton-drag
contribution to these terms thus contains $\partial_i v_j$. Crucially, $\partial_i v_j$ can be captured by operators that appear in the
standard, adiabatic bias expansion, since the Euler equation can be used to relate $\partial_i v_j$ to the tidal field \cite{MSZ}. Thus,
the Compton-drag contributions to the terms of this type are entirely degenerate
with operators appearing in the standard adiabatic bias expansion, and do not need
to be considered further. In fact, the linear Compton-drag term $\propto\theta$ considered above is the simplest example of this.

To summarize, the additional terms induced in the galaxy bias expansion
by Compton drag are up to third order given by\footnote{Unlike the primordial relative velocity $v_{bc}$, which is set in the initial conditions and is thus to be evaluated at the Lagrangian coordinate corresponding to the given Eulerian position, the Compton drag involves the velocity of the matter fluid which is governed by the Euler equation. Hence, there are no displacement terms such as those present for $v_{bc}$ \cite{blazek/etal:15,schmidt:2016b} or in the case of primordial non-Gaussianity \cite{angulo/etal:2015,assassi/baumann/schmidt}.}
\bea
1^{\rm st}\  && - \vs
2^{\rm nd}\  && v^2,\, \v{v}\cdot\v{v}_{bc} \vs
3^{\rm d}\  && v^2\d,\, (\v{v}\cdot\v{v}_{bc}) \d ,\, K_{ij} v^i v^j,\,
K_{ij} v^i v_{bc}^j \,,
\label{eq:list}
\eea
where the tidal field $K_{ij}$ is defined through $K_{ij} \equiv (\partial_i\partial_j/\nabla^2 - \d_{ij}/3) \d$.  The second-order terms contribute to the bias expansion \refeq{biasgen} through
\be
\d_g^{(2)}\Big|_\text{drag} = b_\text{drag} v^2 + b_\text{drag.bc} \v{v}\cdot\v{v}_{bc}\,,
\label{eq:dgbc}
\ee
and appear both in the galaxy three-point function (or bispectrum) and the leading nonlinear (1-loop) correction to the galaxy power spectrum. The third-order terms do not contribute to the galaxy power spectrum at this order, since their contributions are renormalised into lower order bias parameters \cite{mcdonald:2006,assassi/etal,schmidt:2016b}. We will thus focus on the quadratic terms in \refeq{dgbc} in the following. As discussed in \cite{schmidt:2016b,biasreview}, there is also an independent stochastic term associated with each of these operators. These do not appear at the level of the 1-loop galaxy power spectrum and tree-level bispectrum however. In the next section, we will present constraints on the bias parameters $b_\text{drag},\,b_\text{drag.bc}$ from the BOSS DR12 sample.

First, let us provide a very rough estimate for the magnitude of these bias coefficients. Ref.~\cite{dalal/etal:2010} argued, in the context of the primordial contribution $v_{bc}$, that the baryon-CDM relative velocity modulates the local effective sound speed of the gas through
\be
c_{s,\rm eff}^2 = c_s^2|_{v_r=0} + v_r^2\,.
\ee
Using an excursion-set argument for the fraction of gas collapsed into halos, they estimated that the fractional modulation of the galaxy density scales as $v_r^2/c_s^{2}$. Note that $c_s^2$ is, by definition, of the same order of magnitude as the virial velocity of halos with mass $M_{\rm J}$. Hence, this is expected to be a reasonable estimate at least for low-mass halos which are most affected by the relative velocity between baryons and CDM.  We thus obtain
\ba
|b_\text{drag} v^2 | \sim\:& \frac{v_r^2|_\text{drag}}{c_s^2} \sim 5 \times 10^{-8} \frac{v^2}{c_s^{2}} \,,\ \mbox{and hence} \vs
b_\text{drag} \sim\:& 5 \times 10^{-8} c_s^{-2} \sim 30\,,
\label{eq:bdragest}
\ea
where in the first line we have evaluated \refeq{vbcdragest} at $z \sim 6-7$, corresponding to the maximum expected value, and inserted the sound speed for an ideal monatomic gas at $T=10^4\,{\rm K}$. Note in the last relation that we set the speed of light $c=1$. We emphasize that \refeq{bdragest} is only a rough estimate for galaxies formed during or soon after reionization in low-mass halos ($M_h \sim M_{\rm J}$). Thus, $b_\text{drag}$ could be substantially smaller, in particular for galaxies residing in more massive halos. On the other hand, tracers of diffuse gas such as the Ly$\alpha$ forest could be affected more strongly.

Finally, we can also estimate the coefficient $b_\text{drag.bc}$ of the cross-term. If the physical processes leading to the dependence on the primordial baryon-CDM relative velocity and the Compton-drag contribution are fully correlated, i.e. they have the same dependence on the local environment of galaxies, then one expects
\be
|b_\text{drag.bc}| \sim \sqrt{|b_{v^2}^{bc} b_\text{drag}|} \sim 0.6\, \left(\frac{|b_{v^2}^{bc}|}{0.01}\right)^{1/2}\,,
\label{eq:bmixedest}
\ee
where we have used \refeq{bdragest} and assumed that $b_\text{drag.bc}$ as well as $b_{v^2}^{bc}$ refer to the normalized primordial relative velocity $\v{v}_{bc}/\<(v_{bc})^2\>^{1/2}$ [see \refeq{Tthetanorm}], as adopted in \cite{tseliakhovich/hirata:2010,dalal/etal:2010,yoo/dalal/seljak,slepian/etal:2016,beutler/seljak/vlah}. The estimate in \refeq{bmixedest} can be understood as essentially an upper limit. First, the value $b_{v^2}^{bc} \sim 0.01$ is approximately the current upper limit obtained from the BOSS DR12 sample \cite{slepian/etal:2016,beutler/seljak/vlah}. Second, if the physical processes leading to the modulation of the galaxy density by the primordial relative velocity and Compton drag are not directly related, then the amplitude of modulation of the mixed term $\v{v}\cdot\v{v}_{bc}$ is expected to be much smaller than the individual quadratic contributions $v^2, (v_{bc})^2$. This could well be the case, since the effect of the primordial relative velocity is expected to be strongest before reionization, while Compton drag only appears after the onset of reionization. 

\subsection{Pressure and radiative transfer}
\label{sec:bias:RT}

As we have shown in \refsec{pres} and \refsec{RT}, respectively, the effects
of pressure forces and radiative transfer on the clustering of galaxies are,
in the context of the perturbative bias expansion, captured by so-called
higher-derivative contributions. The leading such term is 
\be
\d_g(\vx,\tau) \supset b_{\lapl\d}(\tau) \lapl\d(\vx,\tau)\,.
\label{eq:dghderiv}
\ee
Note that $b_{\lapl\d}$ has units of length squared. 
We will report observational constraints on $b_{\lapl\d}$ in the next section. 
In order to provide an order-of-magnitude estimate, it is useful to consider
a set of higher-order terms of the form
\be
\d_g(\vx,\tau) \supset b_{\lapln\d}(\tau) \lapln\d(\vx,\tau)\,.
\ee
Note that this is only a small subset of all higher-derivative terms; there are also terms such as $(\partial_i\d)^2$, as well as others involving derivatives on the tidal field (see Sec.~2.6 in \cite{biasreview}).  However, this subset of terms will suffice in order to illustrate the expected magnitude of higher-derivative terms.

The higher-derivative terms considered in this paper originate from non-gravitational effects which modulate the galaxy density in a finite region of size $\lambda$, through an expansion of the type \refeq{Iionexp}. We thus expect that the bias parameters scale as
\be
b_{\lapln\d} \sim f\: \lambda^{2n}\,,
\label{eq:hderivest}
\ee
where $|f|$ is the overall amplitude of the modulation (note that $f$ could be positive or negative), while $\lambda$ is the size of the region within which the effects act. In case of pressure forces, this length scale is the Jeans length, $\lambda = \lambda_{\rm J}$, while, if one considers the gas itself as tracer, the amplitude $f$ is of order unity. For the radiative-transfer effects discussed in \refsec{RT}, the relevant scale is $\lambda = \lambda_\text{ion}$, the mean free path of ionizing radiation. On the other hand, the amplitude $f$ of the modulation induced by radiative-transfer effects is uncertain. As argued in \refsec{RT}, one expects the amplitude to be suppressed, $f \ll 1$, for galaxies residing in halos with mass much higher than the Jeans mass $M_{\rm J}$. Similarly to \refeq{hderivest}, one expects that the coefficient of $(\partial_i\d)^2$ will scale as
\be
b_{(\nabla\d)^2} \sim f^2 \lambda^2\,,
\ee
and correspondingly for other nonlinear higher-derivative terms. Thus, if several higher-derivative terms can be measured, then the parameters $f$ and $\lambda$ controlling the higher-derivative contributions can be determined independently. At the level of $b_{\lapl\d}$ alone, there exists a degeneracy between $f$ and $\lambda$. In addition, since $b_{\lapl\d}$ can have either sign, there can also be a chance cancelation of different sources of higher-derivative contributions.

\section{Constraints from the BOSS DR12 sample}
\label{sec:constraints}

\begin{table*}[b]
   \begin{center}
      \caption{Fits to the BOSS DR12 combined sample power spectrum multipoles in the low and high redshift bins ($0.2 < z < 0.5$ and $0.5 < z < 0.75$, respectively). The fit includes the monopole and quadrupole between $0.01 < k < 0.15 h^{-1}$Mpc and the hexadecapole between $0.01 < k < 0.10 h^{-1}$Mpc. All errors in this Table are the marginalised $68\%$ confidence levels, except for the error on the new bias parameters $b_{\rm drag}$ and $b_{\rm drag.bc}$, where we show both the $68\%$ and $95\%$ confidence levels. The labels NGC and SGC refer to the North and South Galactic Cap, respectively.}
      \begin{tabular}{lllllllll}
         \hline
         \multicolumn{9}{c}{Fit to the BOSS DR12 data}\\
 & \multicolumn{4}{c}{$0.2 < z < 0.5$} & \multicolumn{4}{c}{$0.5 < z < 0.75$}\\
         \hline
 & max. like. & mean & max. like. & mean & max. like. & mean & max. like. & mean\\
          $\alpha_{\perp}$  & $ 1.009 $ & $ 1.013 \pm 0.027 $ & $1.010$ & $1.012\pm0.026$ & $ 0.985 $ & $ 0.989 \pm 0.025 $ & $0.985$ & $0.987\pm0.025$\\
          $\alpha_{\parallel}$  & $ 1.006 $ & $ 1.008 \pm 0.040 $ & $1.006$ & $1.014\pm0.041$ & $ 0.975 $ & $ 0.978 \pm 0.041 $ & $0.975$ & $0.977\pm0.041$\\
          $f\sigma_8$  & $ 0.475 $ & $ 0.479 \pm 0.059 $ & $0.466$ & $0.473\pm0.058$ & $ 0.419 $ & $ 0.412 \pm 0.044 $ & $0.418$ & $0.412\pm0.045$\\
          $b_{\rm drag}$  & $ 300 $ & $ 400 \pm 2800 (\pm 5600 )$ & --- & --- & $ -52 $ & $ -35\pm1100 (\pm3100) $& --- & ---\\
          $b_{\rm drag.bc}$  & --- & --- & $-30$ & $-24\pm15(^{+70}_{-31})$ & --- & --- & $-3$ & $-2\pm21(^{+64}_{-38})$\\
\hline
          $b_{1}^{\rm NGC}\sigma_8$  & $ 1.358 $ & $ 1.356 \pm 0.046 $ & $1.359$ & $1.355\pm0.040$ & $ 1.250 $ & $ 1.256 \pm 0.041 $ & $1.252$ & $1.262\pm0.041$\\
          $b_{1}^{\rm SGC}\sigma_8$  & $ 1.347 $ & $ 1.349 \pm 0.058 $ & $1.375$ & $1.361^{+0.053}_{-0.043}$ & $ 1.262 $ & $ 1.261 \pm 0.048 $ & $1.258$ & $1.261\pm0.046$\\
          $b^{\rm NGC}_2\sigma_8$  & $ 1.12 $ & $ 1.12\pm0.77 $ & $1.59$ & $1.13\pm0.78$ & $ 3.07 $ & $ 3.03\pm0.50 $ & $3.05$ & $2.93\pm0.55$\\
          $b^{\rm SGC}_2\sigma_8$  & $ 0.4 $ & $ 0.4 \pm 1.0 $ & $1.39$ & $1.23^{+0.96}_{-0.80}$ & $ 0.80 $ & $ 0.92 \pm 0.92 $ & $0.65$ & $0.86\pm0.93$\\
          N$^{\rm NGC}$  & $ -3000 $ & $ -2800^{+2100}_{-1300} $ & $-700$ & $-600\pm1300$ & $ -2100 $ & $ -2100\pm800$ & $-2100$ & $-2000\pm1800$\\
          N$^{\rm SGC}$  & $ -1700 $ & $ -1400^{+3400}_{-2700} $ & $-1500$ & $-1500\pm1600$ & $ -1200 $ & $ -1400\pm1600 $ & $-900$ & $-1000\pm1700$ \\
          $\sigma^{\rm NGC}_v$  & $ 5.82 $ & $ 5.78 \pm 0.70 $ & $5.93$ & $5.93\pm0.70$ & $ 5.22 $ & $ 5.14 \pm 0.76 $ & $5.22$ & $5.12\pm0.76$ \\
          $\sigma^{\rm SGC}_v$  & $ 6.44 $ & $ 6.43 \pm 0.88 $ & $6.71$ & $6.68\pm0.81$ & $ 4.80 $ & $ 4.61 \pm 0.90 $ & $4.76$ & $4.58\pm0.92$ \\
         \hline
         $\frac{\chi^2}{d.o.f.}$ 
& \multicolumn{2}{c}{$\frac{ 80.7 }{ 74 - 12 }$} & \multicolumn{2}{c}{$\frac{ 79.8 }{ 74 - 12 }$} & \multicolumn{2}{c}{$\frac{ 52.5 }{ 74 - 12 }$} & \multicolumn{2}{c}{$\frac{ 52.5 }{ 74 - 12 }$}\\
         \hline
      \end{tabular}
      \label{tab:relvel_boss}
   \end{center}
\end{table*} 

\begin{table*}[b]
   \begin{center}
      \caption{Fits to the BOSS DR12 combined sample power spectrum multipoles in the low and high redshift bins ($0.2 < z < 0.5$ and $0.5 < z < 0.75$, respectively) including the higher order derivative bias $b_{\lapl \delta}$. The fit includes the same scales as reported in \reftab{relvel_boss}. All errors in this Table are the marginalised $68\%$ confidence levels, except for the error on $b_{\lapl \delta}$, where we show both the $68\%$ and $95\%$ confidence levels.}
      \begin{tabular}{lllllllll}
         \hline
         \multicolumn{5}{c}{Fit to the BOSS DR12 data}\\
 & \multicolumn{2}{c}{$0.2 < z < 0.5$} & \multicolumn{2}{c}{$0.5 < z < 0.75$}\\
         \hline
 & max. like. & mean & max. like. & mean\\
          $\alpha_{\perp}$  & $ 1.009 $ & $ 1.012 \pm 0.027 $ & $0.985$ & $0.986\pm0.027$\\
          $\alpha_{\parallel}$  & $ 1.006 $ & $ 1.010 \pm 0.035 $ & $0.975$ & $0.975\pm0.041$\\
          $f\sigma_8$  & $ 0.475 $ & $ 0.481 \pm 0.055 $ & $0.419$ & $0.424\pm0.047$\\
          $b_{\lapl \delta}\  [(\Mpch)^2]$  & $ -0.1 $ & $ -0.17^{+2.1}_{-1.9}(^{+3.9}_{-3.3})$ & $0.2$ & $0.3^{+1.7}_{-1.9}(^{+3.3}_{-3.8})$\\
\hline
          $b_{1}^{\rm NGC}\sigma_8$  & $ 1.358 $ & $ 1.361 \pm 0.041 $ & $1.249$ & $1.247^{+0.040}_{-0.034}$\\
          $b_{1}^{\rm SGC}\sigma_8$  & $ 1.347 $ & $ 1.348 \pm 0.035 $ & $1.262$ & $1.259^{+0.045}_{-0.042}$\\
          $b^{\rm NGC}_2\sigma_8$  & $ 1.1 $ & $ 1.0\pm1.0 $ & $3.1$ & $3.2\pm0.60$\\
          $b^{\rm SGC}_2\sigma_8$  & $ 0.3 $ & $ 0.2 \pm 0.7 $ & $0.80$ & $0.70\pm0.90$\\
          N$^{\rm NGC}$  & $ -2500 $ & $ -2700^{+2200}_{-1800} $ & $-2000$ & $-2200\pm800$\\
          N$^{\rm SGC}$  & $ -1300 $ & $ -1430\pm1500 $ & $-1250$ & $-1330\pm1300$ \\
          $\sigma^{\rm NGC}_v$  & $ 5.82 $ & $ 5.95 \pm 0.67 $ & $5.22$ & $5.31\pm0.81$ \\
          $\sigma^{\rm SGC}_v$  & $ 6.41 $ & $ 6.53 \pm 0.85 $ & $4.81$ & $4.86\pm0.87$ \\
         \hline
         $\frac{\chi^2}{d.o.f.}$ 
& \multicolumn{2}{c}{$\frac{ 80.7 }{ 74 - 12 }$} & \multicolumn{2}{c}{$\frac{ 52.5 }{ 74 - 12 }$}\\
         \hline
      \end{tabular}
      \label{tab:hderiv_boss}
   \end{center}
\end{table*}

We now provide constraints on the amplitude of the new terms discussed in this paper from the BOSS collaboration DR12 sample~\cite{Eisenstein2011:1101.1529v2, Dawson2012:1208.0022v3}. The galaxy power spectrum measurements we use are described in~\cite{Beutler2016:1607.03150v1} and~\cite{Beutler2016:1607.03149v1} and we refer to these papers for details about the measurements. Note that the BOSS sample is highly biased with $b_1\sim 2$~\cite{Beutler2016:1607.03150v1}, indicating that the luminous red galaxies (LRG) targeted in this sample reside in massive halos $M_h \sim 10^{13}M_{\odot}/h$~\cite{White2012:1203.5306v2}. For this reason, following the discussion in \refsec{bias:drag}, we do not expect particularly strong signatures of the Compton drag, radiative transfer, and primordial baryon-CDM relative velocity in this sample. However, if these halos assembled out of or accreted low-mass halos with $M_h \sim M_{\rm J}$ after reionization, the observable properties of the low-redshift LRG could contain a memory of the reionization effects imprinted on these low-mass halos.
Thus, the constraints we provide on these effects can provide interesting clues on the progenitors of the LRG at redshift $z\sim 8-20$.

We separately constrain the two Compton-drag contributions and the higher-derivative bias $b_{\lapl\d}$. We have found that allowing for either Compton-drag contribution to be varied simultaneously with the higher-derivative bias does not significantly worsen the constraint on $b_{\lapl\d}$, but does increase the allowed range of $b_\text{drag}, b_\text{drag.bc}$ significantly. This is because a change in the low-$k$ power induced by the Compton-drag contributions can be compensated by the change in small-scale power mediated by $b_{\lapl\d}$ to yield a scale-independent change to the galaxy power spectrum, which in turn can be absorbed by a shift in $b_1$. This degeneracy is expected to be broken if the galaxy three-point function or bispectrum is included. Further, the data are currently not sufficiently sensitive to distinguish between the Compton-drag and primordial relative-velocity contributions. Thus, we constrain all three contributions individually, and leave meaningful combined constraints using higher-order statistics to future work. Note that the individual constraints allow us to obtain a conservative estimate for possible systematics in the BAO scale induced by the Compton-drag contributions.

\subsection{Compton drag}

\begin{figure}[t]
\begin{center}
  \includegraphics[width=8cm]{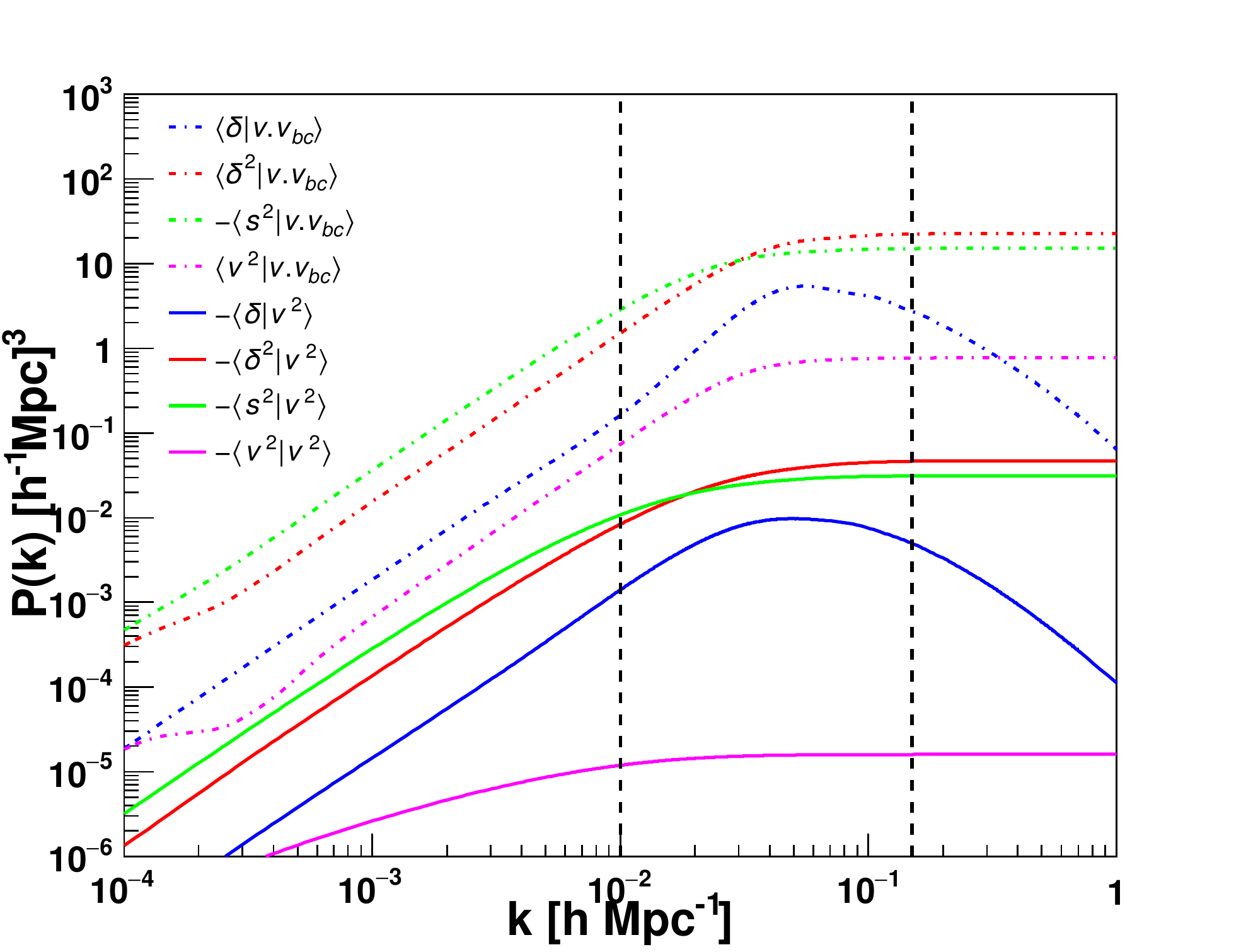}
\caption{Contributions to the 1-loop galaxy power spectrum due to the leading Compton-drag terms [\refeq{dgbc}]: $v^2$ (solid lines) and $v\cdot v_{bc}$ (dashed lines). }
\label{fig:cterms_cmp}
\end{center}
\end{figure}

As we have seen in the previous section, the Compton-drag contribution to the baryon-CDM relative velocity is captured by terms involving the matter velocity $\v{v}$. For the purpose of the galaxy
power spectrum including the leading nonlinear correction, there are two
additional operators to consider, as given in \refeq{dgbc}. 
The structure of the new terms in \refeq{dgbc} is very similar to those of the primordial baryon-CDM
relative velocity $v_{bc}$. Hence, the prediction for the 1-loop galaxy power
spectrum can be taken from Appendix~A of \cite{beutler/seljak/vlah}, 
with the following differences:
\begin{enumerate}
\item The terms involving the relative density perturbation $\d_{bc}$ and divergence of the relative velocity $\theta_{bc}$ are absent. This is formally obtained by setting $b_\d^{bc} = 0 = b_\theta^{bc}$ in the notation of \cite{beutler/seljak/vlah}. Further, the so-called advection terms are absent as well, formally obtained by setting $L_s=0$.
\item The kernel $G_u$, defined in Eq.~(24) of \cite{beutler/seljak/vlah}, is modified to
  \begin{align}
   G_{\rm drag}(k_1,k_2) &= -(\cH f)^2/(k_1 k_2), \\
   G_{\rm drag.bc}(k_1,k_2) &= T_v(k_1) \cH f/k_2\,,
  \end{align}
  for the quadratic Compton-drag term ($\propto b_{\rm drag}$) and the coupling between Compton drag and the primordial relative velocity ($\propto b_{\rm drag.bc}$), respectively. Here, $T_v(k) \equiv T_{\theta_{bc}}(k)/[k T_m(k)]$, where $T_{\theta_{bc}}$ is the transfer function of the primordial relative velocity divergence defined analogously to \refeq{Tdef}, but normalized to unit variance of the relative velocity via \refeq{Tthetanorm} (see also Eq.~(23) and following in \cite{beutler/seljak/vlah}).
\end{enumerate}
For reference, we explicitly give the complete expression for the 1-loop galaxy power spectrum in \refapp{Pkmodel}. The Compton-drag contributions to the galaxy power spectrum are shown in \reffig{cterms_cmp}. Their different scale dependence compare to the terms in the standard bias expansion allows us to place constraints even after marginalizing over all bias parameters.

We compare our model to the BOSS power spectrum monopole, quadrupole and hexadecapole in the wavenumber range $0.01$ - $0.15 \iMpch$ for the monopole and quadrupole and $0.01$ - $0.1\iMpch$ for the hexadecapole. In addtion to our parameters of interest $b_\text{drag}$, $b_\text{drag.bc}$, which we constrain in turn, our fit has 7 free parameters: the BAO and redshift-space distortion scaling parameters $\alpha_{\perp}$, $\alpha_{\parallel}$, $f\sigma_8$; and the $4$ bias and stochastic parameters $b_1$, $b_2$, $\sigma_v$, $N$. For the latter 4 parameters, we allow for independent values for the North Galactic Cap (NGC) and South Galactic Cap (SGC), as these involve slightly different selection functions (see~\cite{Beutler2016:1607.03150v1} for details). Note that the galaxy samples only differ slightly, and we thus expect the values of $b_\text{drag},\,b_\text{drag.bc}$ to be similar. Since we are only able to obtain upper limits on these parameters, we refrain from treating the NGC and SGC samples separately, and constrain a single parameter for the entire sample.

\begin{figure}[t]
\begin{center}
  \includegraphics[width=8cm]{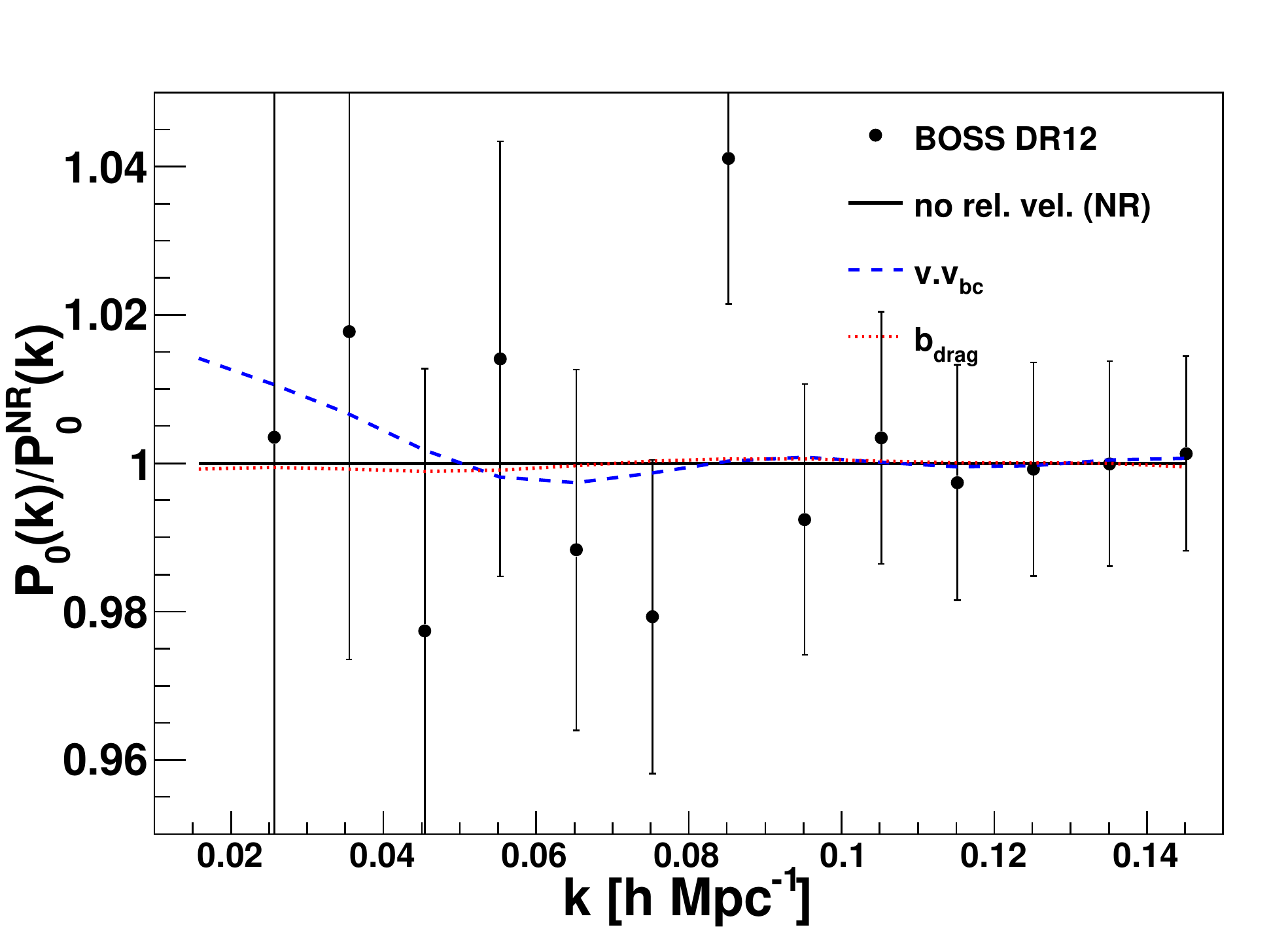}
\caption{The BOSS DR12 monopole power spectrum measurement (black data points) divided by the best fitting model without relative velocity parameters (NR, black solid line). The red dotted line shows the best fitting model when including $b_\text{drag} v^2$,  while the blue dashed line shows the best fitting model when including $b_\text{drag.bc} \vv\cdot \vv_{\rm bc}$. Here we only show the monopole for clarity. While the best fit has been obtained by fitting the monopole, quadrupole and hexadecapole, the constraints on $b_\text{drag}, b_\text{drag.bc}$ are dominated by the monopole.}
\label{fig:cterms_bestfit}
\end{center}
\end{figure}

The monopole of the best-fit model is shown in \reffig{cterms_bestfit}. Note the precision of the data, which constrain the galaxy power spectrum in this range of scales to within a few percent. 

All fitting results are summarised in \reftab{relvel_boss}. Fitting this model to the BOSS power spectrum we obtain constraints of $b_{\rm drag} = 400 \pm 5600$ ($95\%$ confidence level) and $b_{\rm drag} = -35\pm 3100$ for the low and high redshift bin, respectively. We see that the $95\%$ confidence interval is about two orders of magnitude larger than the rough order-of-magnitude estimate of the previous section, $|b_{\rm drag}| \sim 30$ [\refeq{bdragest}]. 

The cross terms between the Compton drag and the primordial relative velocity are constrained to $b_{\rm drag.bc} = -24^{+70}_{-31}$ ($95\%$ confidence level) and $b_{\rm drag.bc} = -2^{+64}_{-38}$ for the low and high redshift bin of BOSS, respectively. Again, the observational uncertainties are significantly larger than the maximum expected value of this bias parameter, $|b_\text{drag.bc}| \lesssim 0.6$ [\refeq{bmixedest}].

Thus, the data show no evidence for the presence of modulations due to Compton drag and primordial relative velocity, which is expected given the size of the errors. We will discuss possibilities for future improvements on these constraints in \refsec{concl}.

\subsection{Radiative transfer and pressure}

We now turn to constraints on the higher-derivative terms introduced in the galaxy bias expansion by radiative-transfer and pressure effects. Unfortunately, higher-derivative contributions are difficult to constrain from the galaxy power spectrum, as they are partially degenerate with higher-order nonlinear contributions that are not included in the model. For this reason, we only attempt to constrain the leading higher-derivative bias here, $b_{\lapl\d}$ in \refeq{dghderiv}. This simply adds a single term to the galaxy power spectrum,
\ba
P_{gg}(k,\mu)\Big|_\text{h. deriv.} =\:& -2 (b_1 + f\mu^2) b_{\lapl\d} k^2 P_m(k)\,.
\ea
Recall that $b_{\lapl\d}$ has dimension length squared. We have not included the contribution $(b_{\lapl\d})^2 k^4 P_m(k)$, as it is of the same order as higher nonlinear contributions which we do not include in our model.  
  Given the very different scale dependence compared to the Compton-drag contributions, it is justified to set the latter to zero when constraining $b_{\lapl\d}$. The remaining parameters are allowed to be free, as described above. The results are given in \reftab{hderiv_boss}.  At the 95\% confidence level, we obtain approximately $|b_{\lapl\d}| \lesssim (2\Mpch)^2$. Interestingly, this is of the same order as what one would expect for the higher-derivative bias of the parent halos of the LRG sample; the Lagrangian radius of $M_h= 10^{13}\Msunh$ halos is $2\Mpch$.

Turning to the interpretation of this result, there is a degeneracy between the amplitude and spatial length scale of the pressure or radiative-transfer modulation, as described in \refsec{bias:RT}. However, given an assumption on the relevant mean-free path of ionizing radiation, we can turn the constraint on $b_{\lapl\d}$ into a constraint on the amplitude of the modulation by the ionizing background, via \refeq{hderivest}:
\be
|f| \lesssim 0.002 \left(\frac{\lambda_\text{ion}}{50 \Mpch}\right)^{-2}\,.
\ee
We see that, for the BOSS DR12 galaxy sample, a modulation within a region of comoving size $50 \Mpch$, say due to radiative-transfer effects, has to be of very small amplitude. Of course, in case of a modulation on much smaller spatial scales, the constraint on the amplitude becomes significantly weaker.

\section{Summary and conclusions}
\label{sec:concl}

Reionization is the first and only phase transition in the Universe since
recombination, 400,000 years after the Big Bang, and clearly a key event
in the history of baryonic matter in the Universe. In this paper, 
we have investigated the impact of reionization on the large-scale clustering
of galaxies, in the context of the perturbative bias expansion (see \cite{biasreview} for a review). We have identified three distinct contributions. The first two, \bfem{Compton drag} and \bfem{pressure}, are due to non-gravitational forces acting on baryons and leading to a relative velocity (and subsequently density) of the baryons with respect to CDM. These two contributions, together with the primordial contribution induced by pre-recombination plasma oscillations, completely describe the scale- and redshift-dependence of the baryon-CDM relative velocity at linear order. 
The third, \bfem{radiative transfer}, is due to the modulation of the heating and cooling rates of the gas by the local background of ionizing radiation. We have systematically derived the complete contributions of these terms to the nonlinear galaxy bias expansion, and constrained the leading contributions using the BOSS DR12 sample.

The \bfem{Compton drag} due to the motion of the gas relative to the CMB leads to unique contributions to the galaxy density, which involve the velocity of matter directly. In the absence of non-gravitational forces, this term is forbidden by the equivalence principle. However, the CMB corresponds to a locally identifiable preferred frame, with respect to which the velocity is defined. 
The leading Compton-drag term is $b_{\rm drag} v^2$, whose coefficient $b_{\rm drag}$ we have constrained using the BOSS DR12 sample. The constraint,
$|b_{\rm drag}| \lesssim 3000$ at 95\% CL for the high-redshift sample and a factor of 2 worse for the low-redshift sample, is a factor of $\sim 100$ larger than the value expected from a rough order-of-magnitude forecast for objects of order the post-reionization Jeans mass. We have also constrained the mixed contribution involving both Compton drag and primordial relative velocity. Again, the constraint is several orders of magnitude above what is expected from a rough estimate. While the constraints on the Compton-drag contributions appear to be much weaker than those on the primordial relative velocity,
$|b_{v^2}^{bc}| \lesssim 0.01$ \cite{slepian/etal:2016,beutler/seljak/vlah}, this is merely due to the essentially arbitrary normalization chosen for $v_{bc}$ ($\< v_{bc}^2 \> =1$ at $z=0$). If we normalize $v_{bc}$ to the speed of light, as done with the Compton-drag contribution here, then the constraints translate to $|b_{v^2}^{bc}| \lesssim 0.01 c^2 / \< v^2_{bc}\> \approx 10^{12}$ at $z=0$.
We emphasize that, given our constraints, the broad-band amplitude of the Compton-drag contribution to the galaxy power spectrum is small, and higher-order Compton-drag effects will be even smaller. Hence, they do not change the reach of perturbative appraoches to galaxy statistics.

Since these contributions have never been considered before, it is also interesting to investigate whether the Compton-drag contributions could shift the BAO feature in the galaxy two-point function which is used as a standard ruler in cosmology. Given our $95\%$ confidence-level constraints, the potential fractional shifts in the transverse and radial BAO scales are found to be limited to $0.01\%$ and $0.1\%$ for $b_{\rm drag}$ and $b_{\rm drag.bc}$, respectively. On the other hand, the potential bias in the measured growth rate parameter $f\sigma_8$ is $0.5\%$ and $3\%$ for $b_{\rm drag}$ and $b_{\rm drag.bc}$, respectively. The stronger parameter shifts obtained for $b_{\rm drag.bc}$ are presumably due to its scale dependence (cf. \reffig{cterms_cmp}) which is less degenerate with other nonlinear contributions to the galaxy power spectrum and thus not as easily absorbed by the standard bias parameters. Still, the parameter shifts are well below $0.5\sigma$ of the uncertainties of current BOSS DR12 measurements. We stress however that the potential biases in the growth rate are not negligible for next-generation galaxy redshift surveys such as HETDEX, DESI, Euclid, PFS, and WFIRST.

The \bfem{pressure and radiative-transfer} effects are captured by higher-derivative terms. We have constrained the leading representative, $b_{\lapl\d} \lapl\d$, to a value of $|b_{\lapl\d}| \lesssim (2\Mpch)^2$. This tightly constrains any large-scale ($\gtrsim 20 \Mpch$) modulation by radiative-transfer effects of the progenitors of the BOSS DR12 sample. However, modulations on smaller spatial scales are much less constrained. In order to disentangle the amplitude and spatial scale of modulations to radiative-transfer effects or pressure forces, subleading higher-derivative contributions also need to be measured (\refsec{bias:RT}). This is clearly worthwhile, as perturbative approaches to galaxy clustering break down on the spatial scale of the modulation.

As is generally true for bias parameters beyond the linear bias $b_1$, we expect significant improvement in the constraints on $b_{\rm drag},\,b_{\rm drag.bc},\,b_{\lapl\d}$ when combining the galaxy power spectrum with the galaxy three-point function or bispectrum. For the leading, tree-level three-point function, this can immediately be done using the terms given in \refsec{bias}. We leave this for future work.

\acknowledgments
We thank Koki~Kakiichi and Simon~White for helpful discussions. 
FS acknowledges support from the Marie Curie Career Integration Grant  (FP7-PEOPLE-2013-CIG) ``FundPhysicsAndLSS,'' and Starting Grant (ERC-2015-STG 678652) ``GrInflaGal'' from the European Research Council. FB acknowledges support from the UK Space Agency through grant ST/N00180X/1.

\begin{widetext}
\appendix

\section{The power spectrum model}
\label{app:Pkmodel}

The galaxy power spectrum in redshift-space at 1-loop order, which we use to describe the BOSS galaxy power spectrum measurements, was derived in~\citep{McDonald2009:0902.0991v1, Taruya2010:1006.0699v1, Saito2014:1405.1447v4} and has been extensively tested in~\citep{Beutler2013:1312.4611v2, Beutler2016:1607.03150v1, beutler/seljak/vlah}.  This prediction is given by
\begin{equation}
\begin{split}
P_{\rm g, NL}(k,\mu) &= \exp\left\{-(fk\mu\sigma_v)^2\right\}\left[P_{{\rm g},\delta\delta}(k) + 2f\mu^2P_{{\rm g},\delta\theta}(k) + f^2\mu^4P_{\theta\theta}(k) + b_1^3A(k,\mu,\beta) + b_1^4B(k,\mu,\beta)\right], 
\label{eq:taruya}
\end{split}
\end{equation}
where $\mu = k_z/k$ is the cosine of the wavenumber with the line of sight, which we choose along the $z$ axis.
For more details on the definition of the individual terms we refer to~\citep{Beutler2013:1312.4611v2, Beutler2016:1607.03150v1} were this model is explained in detail. This prediction, however, neglects all contributions from baryon-CDM relative velocity and density perturbations. 
We now add all relevant contributions of these latter effects. In the following, we will let $\d$ and $\theta$ stand for the total matter density perturbation and velocity divergence, respectively.

In order to write the equations in a compact way, we let $O,Q,\dots$ stand for operators such as density, velocity squared, and so on, and $b_O, b_Q, \dots$ for the corresponding bias parameters.  We group the operators appearing here into three sets: $\B_m$, corresponding to the terms appearing in the standard bias expansion for adiabatic initial conditions, i.e. without relative baryon-CDM perturbations; $\B_{bc}$, the operators induced by the primordial relative density and velocity perturbations, which were considered in \cite{dalal/etal:2010,yoo/dalal/seljak,tseliakhovich/barkana/hirata,Visbal/etal:12,popa/etal,yoo/seljak,bernardeau/vdr/vernizzi,slepian/eisenstein,lewandowski/perko/senatore,blazek/etal:15,schmidt:2016b,ahn}; and $\B_\text{drag}$, the contributions from Compton drag which were first derived here.  We further split these sets into first (e.g., $\B_m^{(1)}$) and second order (e.g., $\B_m^{(2)})$ in perturbations. As is done in \cite{beutler/seljak/vlah}, we simplify the result by neglecting the quadratic operators involving the primordial baryon-CDM relative density perturbation $\delta_{bc}$, and include $v_{bc}^2$ as only quadratic term involving the primordial relative velocity. These can be straightforwardly added following \cite{schmidt:2016b}. In any case, they are not considered for the results in this paper. When making this simplifying assumption, no new cubic-order bias terms appear in the 1-loop power spectrum, and the single cubic term is included in \refeq{taruya} above. Further, we do not explicitly write the advection term of \cite{blazek/etal:15}, as it can be absorbed in the bias coefficient of $\theta_{bc}$ \cite{schmidt:2016b}. 

We then have
\ba
\B_m^{(1)} \equiv\:& \{ \d \}\,, \quad \B_m^{(2)}\equiv \{ \d^{(2)},\,\d^2,\,s^2 \}\,,\quad\mbox{where}\quad
s^2 \equiv \left[\left(\frac{\partial_i\partial_j}{\lapl} - \frac13 \d_{ij}\right)\d\right]^2  \vs
\B_{bc}^{(1)} \equiv\:& \{ \d_{bc},\,\theta_{bc} \}\,,\quad
\B_{bc}^{(2)} \equiv \{ v_{bc}^2 \} \vs
\B_\text{drag}^{(1)} \equiv\:& \{\  \}\,, \quad
\B_\text{drag}^{(2)} \equiv \{ v^2,\, \v{v}\cdot\v{v}_{bc} \}\,,
\ea
Here, $\d^{(2)}$ denotes the second-order density field in standard perturbation theory, while operators without superscripts are assumed to be constructed out of linear fields (i.e., $\d^2 \equiv (\d^{(1)})^2$, and so on).
Further, $b_{\d^{(2)}} = b_\d = b_1$, while $b_{\d^2} = b_2/2$, and we have denoted $b_\text{drag} \equiv b_{v^2}$, $b_\text{drag.bc} \equiv b_{\vv\cdot\vv_{bc}}$, and $b_{v^2}^{bc} \equiv b_{v_{bc}^2}$ in the text for clarity.  
For convenience, we also define $\B_\text{all}^{(i)} \equiv \B_m^{(i)} \cup \B_{bc}^{(i)} \cup \B_\text{drag}^{(i)}$.  

Then, the full expression for the 1-loop galaxy power spectrum in redshift space becomes
\ba
P_g(k,\mu) =\:& P_{\rm g, NL}(k,\mu) + \sum_{O \in \B_{bc}^{(1)} \cup B_\text{drag}^{(1)}} \left[ \sum_{Q\in \B_\text{all}^{(1)}} b_O b_Q P_{O|Q}^{(11)}(k) - 2  \mu^2 P_{O|\hat\theta}^{(11)}(k) \right]
\vs
& +\sum_{O \in \B_{bc}^{(2)} \cup \B_\text{drag}^{(2)}} b_O \left[ \sum_{Q \in \B_\text{all}^{(2)}} b_Q P_{O|Q}^{(22)}(k) - 2  \mu^2 P_{O|\hat\theta^{(2)}}^{(22)}(k) -2 P_{O|\d \hat\eta}^{(22)}(k,\mu) - P_{O|\hat\eta^2}(k,\mu) \right]
\label{eq:psmodel}
\ea
where $\hat\theta \equiv \vn\cdot\v{v}/\cH$ is the scaled matter velocity divergence, while $\hat\eta \equiv \hat n^i \hat n^j \partial_i v_j/\cH$ is the line-of-sight derivative of the line-of-sight velocity.  In Fourier space, $\hat\eta(\vk) = \mu^2 \hat\theta(\vk)$ and, at linear order, $\hat\theta = - f \d$.
Here, the linear terms are given by 
\ba
\< O^{(1)}(\vk) Q^{(1)}(\vk')\> = P_{O|Q}^{(11)}(k) (2\pi)^3 \d_D(\vk+\vk')\,.
\ea
For example, we have
\be
P_{\d_{bc}|\d}^{(11)}(k) = \frac{T_{\d_{bc}}(k)}{T_m(k)} P_m^{\rm lin}(k)
\quad\mbox{and}\quad
P_{\d_{bc}|\hat\theta}^{(11)}(k) = - f \frac{T_{\d_{bc}}(k)}{T_m(k)} P_m^{\rm lin}(k)\,,
\ee
where the transfer functions for the primordial relative density perturbation $\d_{bc}$ and relative velocity divergence $\theta_{bc}$ are defined in analogy to $T_{\theta_{r}}(k)$ through \refeq{Tdef}, except that we normalize $T_{\theta_{bc}}$ so that
\be
\< \vv_{bc}^2\> = \int \frac{d^3\vk}{(2\pi)^3} k^{-2} \left(\frac{T_{\theta_{bc}}(k)}{T_m(k)}\right)^2 P_m^{\rm lin}(k) = 1 \quad\mbox{at}\  z=0\,.
\label{eq:Tthetanorm}
\ee
As discussed in \refsec{bias:drag}, in practice the transfer function for $\theta_{bc}$ can be obtained from a pre-reionization output of the relative-velocity tranfer function $T_{\theta_r}$ computed by a Boltzmann code, since the Compton-drag term is absent before reionization.

\begin{table}[b]
\centering
\begin{tabular}{l|l}
\hline
\hline
Operator $O$ & Kernel $G_O(\vk_1,\vk_2)$, with $\mu_{12} \equiv \vk_1\cdot\vk_2/(k_1 k_2)$ \\
\hline
$\d^{(2)}$ & $F_2(\vk_1,\vk_2)$ \\
$\hat\theta^{(2)}$ & $- f G_2(\vk_1,\vk_2)$ \\
$\d^2$ & 1 \\
$s^2$ & $\mu_{12}^2 - 1/3$ \\
$v_{bc}^2$ & $- \mu_{12} T_{\theta_{bc}}(k_1) T_{\theta_{bc}}(k_2) / [k_1 T_m(k_1) k_2 T_m(k_2)]$ \\
$v^2$ & $- \mu_{12} \cH^2 f^2/(k_1k_2)$ \\
$\v{v}\cdot\v{v}_{bc}$ & $\mu_{12} \cH f T_{\theta_{bc}}(k_1)/[k_1 T_m(k_1) k_2]$ \\
\hline
\hline
\end{tabular}
\caption{List of second-order operators and corresponding kernels appearing in the contributions $P_{O|Q}^{(22)}(k)$ [\refeq{P22} via \refeq{GOs}] to the 1-loop galaxy power spectrum. $F_2$ and $G_2$ are the second order density and velocity kernels, respectively (e.g., \cite{Bernardeau/etal:2002}).
\label{tab:kernels}}
\end{table}

With the exception of the last two terms, the nonlinear terms on the second line of \refeq{psmodel} can be written as
\ba
P_{O|Q}^{(22)}(k) \equiv 2 \int \frac{d^3\vq}{(2\pi)^3} P_m^{\rm lin}(k) P_m^{\rm lin}(|\vk-\vq|) \left[G_O^s(\vq, \vk-\vq) G_Q^s(\vq,\vk-\vq) - G_O^s(\vq,-\vq) G_Q^s(\vq,-\vq) \right]\,,
\label{eq:P22}
\ea
where the symmetrized kernels are defined as
\be
G_O^s(\vk_1,\vk_2) \equiv \frac12 \left[ G_O(\vk_1,\vk_2) + G_O(\vk_2,\vk_1)\right]\,,
\label{eq:GOs}
\ee
and the kernels $G_O$ are listed for each operator in \reftab{kernels}.
Finally, the last two terms in \refeq{psmodel} are given by
\ba
P_{O|\d\hat\eta}(k,\mu) =\:& - f\int \frac{d^3 q}{(2\pi)^3} G_O^s(\vq,\vk-\vq) 
\left[\frac{(\vnhat\cdot\vq)^2}{q^2} + \frac{[\vnhat\cdot(\vk-\vq)]^2}{|\vk-\vq|^2}
  \right]
P_m^{\rm lin}(q) P_m^{\rm lin}(|\vk-\vq|)
\vs
P_{O|\hat\eta^2}(k,\mu) =\:& 2f^2 \int \frac{d^3 q}{(2\pi)^3} G_O^s(\vq,\vk-\vq) 
\frac{(\vnhat\cdot\vq)^2}{q^2} \frac{[\vnhat\cdot(\vk-\vq)]^2}{|\vk-\vq|^2} P_m^{\rm lin}(q) P_m^{\rm lin}(|\vk-\vq|)\,.
\ea
Note that the absence of preferred directions apart from the line of sight $\vnhat$ ensures that these correlators only depend on $k$ and its angle $\mu$ with $\vnhat$.

It is worth noting that \refeq{psmodel} includes terms such as $P_{O|\d\hat\eta}(k)$ that couple nonlinear redshift-space distortions with nonlinear relative-velocity-induced bias terms. There are corresponding terms for the standard nonlinear bias operators $\d^2,\, s^2$ that are not included in the model of \citep{Taruya2010:1006.0699v1, Saito2014:1405.1447v4}, which therefore do not correspond to the complete expression for the 1-loop galaxy power spectrum in redshift space. However, we do not expect these missing terms to have any impact on the constraints on the Compton drag and primordial relative velocity contributions. Note that \cite{beutler/seljak/vlah} found that these constraints are dominated by the power spectrum monopole.

\end{widetext}

\section{The adiabatic decaying mode}
\label{app:decaying}

\begin{figure*}[t]
\centering
\includegraphics[width=0.9\textwidth]{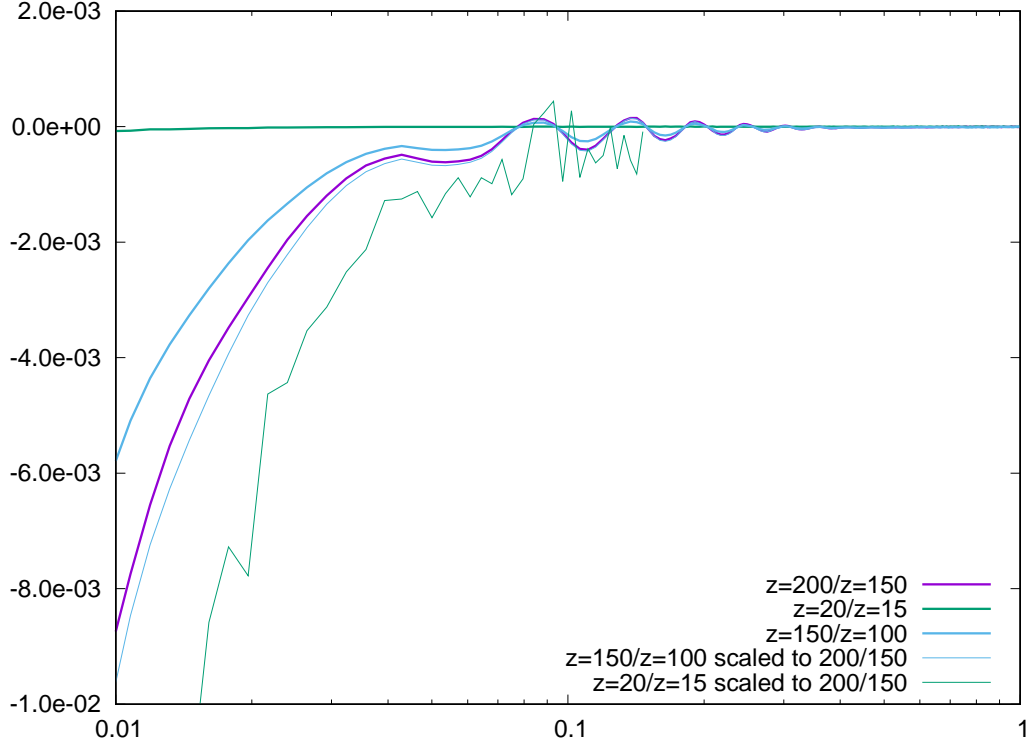}
\caption{Transfer function of the adiabatic decaying mode, determined using \refeq{Tmratio} for different values of $(z_1,z_2)$.  
\label{fig:Tkdec}}
\end{figure*}

Since the local abundance of galaxies in general depends on the entire
history of structure formation, all modes of the cosmic density fields
(in particular baryons $b$ and CDM $c$) should be included. One such mode has
so far been neglected in the large-scale structure literature, namely the adiabatic decaying mode, whose time dependence is given by $H(z)$.  Here, adiabatic means that $\d_i = \d_m$, $i=b,c$.
In order to estimate the magnitude of this mode, we write the
redshift-dependent matter transfer function as
\be
T_m(k,z) = T_+(k) \hat D(z) + T_-(k) E(z)\,,
\label{eq:Tm}
\ee
where $T_m = f_b T_b + (1-f_v) T_c$, and $T_\pm$ are the transfer functions of the adiabatic growing and
decaying modes normalized to $z=0$, $\hat D(z)$ is the normalized growth
factor with $\hat D(0)=1$, and $E(z) \equiv H(z)/H_0$.  
Hence,
\be
\frac{\hat D(z_2) T_m(k,z_1)}{\hat D(z_1) T_m(k,z_2)}-1 = \frac{T_-(k)}{T_+(k)}
\left( \frac{E(z_1)}{\hat D(z_1)} - \frac{E(z_2)}{\hat D(z_2)} \right)\,,
\label{eq:Tmratio}
\ee
which allows us to estimate the amplitude of the decaying mode relative
to the growing mode from transfer function outputs at two redshifts.
\reffig{Tkdec} shows the r.h.s. of \refeq{Tmratio} for differenct combinations of redshifts (with $z_1>z_2$). Note that \refeq{Tm} neglects the gravitational
coupling to all other stress-energy components, in particular radiation and
neutrinos. Hence, we do not expect it to exactly describe the transfer function
given by Boltzmann codes, especially on scales approaching the comoving horizon
where radiation and neutrino contributions are most significant.
Given these caveats, we limit \reffig{Tkdec} to scales of $k \gtrsim 0.01 \iMpch$.  
In order to verify that the result actually corresponds to the desired decaying mode, we also show the result of rescaling the different results
to a common redshift ratio of $(z_1=200, z_2=150)$ using the expected behavior of the decaying mode (assuming matter domination). 
The approximate match confirms that the adiabatic decaying mode does correspond to the bulk of \refeq{Tmratio}. 

We see that the decaying mode is strongly suppressed relative to the
growing mode already at $z\simeq 150$.  Quantitatively, we obtain
\be
\frac{T_-(k) E(z)}{T_+(k) \hat D(z)} \lesssim 6\times 10^{-8} \frac{E(z)}{\hat D(z)}
\sim 6\times 10^{-8} \left(1+z\right)^{5/2}\,,
\ee
where the second approximate equality assumes matter domination.
Even at redshift 10, this is only of order $10^{-5}$. Hence, including the
decaying mode in the bias expansion, with a bias coefficient of order one,
yields a contribution which is
entirely negligible for current and upcoming galaxy surveys.

\bibliography{REFS}

\end{document}